%% file: main.tex
\documentclass[twocolumn,journal]{IEEEtran}
\usepackage{color}
\input{my_sections.tex}

\input{mysymbol.sty}
\usepackage{theorem}
\usepackage{cite}
\usepackage{amsmath}
\usepackage{amssymb}
\usepackage{mathtools}
\usepackage{multirow}
\usepackage{booktabs}
\usepackage{enumitem}
\usepackage{url}
\usepackage{graphics, subfigure, times, amsfonts}
\usepackage{tikz, epic,eepic}
\usetikzlibrary{shapes,arrows}
\usepackage{pgfplots}
\usepackage{color}
\usepackage{hyperref}

\usepackage{srcltx} 
\usepackage{latexsym}
\usepackage{amscd, verbatim}

\newcommand{\QED}{\hfill\ensuremath{\blacksquare}}
\newtheorem{theorem}{Theorem}
\newtheorem{claim}{Claim}
\newtheorem{algorithm}{Algorithm}
\newtheorem{proposition}{Proposition}
\newtheorem{corollary}{Corollary}
\newtheorem{lemma}{Lemma}
\newtheorem{definition}{Definition}
\newtheorem{example}{Example}
{\itshape}{\rmfamily}

\newtheorem{remark}{Remark}

\def\forall{\text{for all\ }}

\mathtoolsset{showonlyrefs}

\title{Control of learning in anti-coordination network games}
\author{Ceyhun Eksin$^\dagger$ and Keith Paarporn$^\ddagger$
\thanks{$^\dagger$Industrial and Systems Engineering Department, Texas A\&M University, College Station, TX. {\tt\small eksinc@tamu.edu}}
\thanks{$^\ddagger$Electrical and Computer Engineering, University of California, Santa Barbara, CA. {\tt\small kpaarporn@ucsb.edu}}}

\begin{document}
\normalsize
\maketitle

%
\begin{abstract}
We consider control of heterogeneous players repeatedly playing an anti-coordination network game. In an anti-coordination game, each player has an incentive to differentiate its action from its neighbors. At each round of play,  players take actions according to a learning algorithm that mimics the iterated elimination of strictly dominated strategies. We show that the learning dynamics may fail to reach anti-coordination in certain scenarios. We formulate an optimization problem with the objective to reach maximum anti-coordination while minimizing the number of players to control. We consider both static and dynamic control policy formulations. Relating the problem to a minimum vertex cover problem on bipartite networks, we develop a feasible dynamic policy that is efficient to compute. Solving for optimal policies on benchmark networks show that the vertex cover based policy can be a loose upper bound when there is a potential to make use of cascades caused by the learning dynamics of uncontrolled players. We propose an algorithm that finds feasible, though possibly suboptimal, policies by sequentially adding players to control considering their cascade potential. Numerical experiments on random networks show the cascade-based algorithm can lower the control effort significantly compared to simpler control schemes.
\end{abstract}

\section{Introduction}

\input{introduction.tex}

\section{Anti-coordination network games}\label{sec:model}
\begin{figure}[t]
	\centering
	\input{fig_1.tex}
	\caption{\small Network game with binary types. }
	\label{fig:bipartite}
\end{figure}
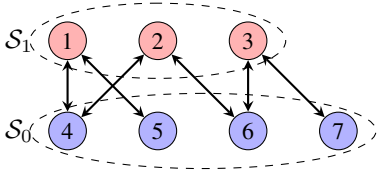

A game consists of a population of players $i \in \ccalN:=\{1,\dots, n\}$ that take action $a_i \in \ccalA_i$ in order to maximize their utility function $u_i(\cdot)$. We assume each player is in one of two possible types $s_i\in\{0,1\}$. The population is divided into two disjoint sets $\ccalS_0:=\{i \in \ccalN: s_i =0 \}$ and $\ccalS_1:=\{i \in \ccalN : s_i =1 \}$. Only the actions of neighbors that have the opposite type can affect a player's utility function. For instance, if a player is type 0, then its utility depends on actions of its neighbors in $\ccalS_1$. We can capture the payoff dependency of players using a bipartite graph $\ccalG_B = (\ccalS_0, \ccalS_1, \ccalE_B)$---see Figure \ref{fig:bipartite}. We define the neighborhood of player $i$ as $\ccalN_i := \{j\in \ccalN: (i,j) \in \ccalE_B\}$. Then the utility function is defined as $u_i: \ccalA_i \times \ccalA_{\ccalN_i} \to \reals$ where $\ccalA_{\ccalN_i} = \prod_{j\in \ccalN_i} \ccalA_j$. 
Given the bipartite network $\ccalG_B$, the network game with binary types can be represented by the tuple $\Gamma=\{\ccalN, \ccalA, \ccalG_B, \{u_i\}_{i\in \ccalN}\}$.

The premise of an anti-coordination game is that a player benefits if its opponents yield. Similarly, in an anti-coordination network game, a player benefits if its neighbors in the opposing type yield. We assume player $i$ can take actions between zero and one, i.e., $\ccalA_i = [0,1]$ for all $i \in \ccalN$. The following utility function, 
\begin{align} \label{utility}
 u_i(a_i,  &a_{\ccalN_i})=  a_i \left(1 -  \big(c_0 (1-s_i) +  c_1 s_i \big)\sum_{j\in \ccalN_i} a_j \right)
\end{align}
with $c_0 \in (0,1)$ and $c_1\in (0,1)$ as constants, captures the preferences of players to differentiate their actions from their neighbors. Here, $a_{\ccalN_i}$ are the actions of player $i$'s neighbors. Action $a_i = 1$ maximizes the utility if the term inside the parentheses is positive. Otherwise, action $a_i = 0$ maximizes the utility. The constant 1 inside the parentheses means that the preferred action is 1 regardless of the type of the player $i$. The term that is subtracted from one captures the decrease in the preference of the player to choose action 1. That is, as $i$'s neighbors increase their action, the benefit of $i$ from choosing the preferred action decreases. This decrease depends on the type of the player. If the player's type is 0 (1), i.e., $s_i=0$ $(=1)$, then the decrease is proportional to $c_0$ $(c_1)$. 

Below we provide examples for the anti-coordination network game $\Gamma$ with payoffs as in \eqref{utility}. 

\begin{example}[Disease spread on networks]
Players want to avoid disease transmission \cite{eksin2017disease}. Each player is either healthy ($s_i = 0$) or sick ($s_i = 1$). The network $\ccalG_B$ is a contact network with each edge representing a chance of disease transmission between a healthy and a sick player. The action space captures the social distancing level of a player with action $a_i = 0$ representing self-isolation and action $a_i = 1$ representing resuming normal activity. Actions between 0 and 1 represent different levels of disease prevention measures, e.g., covering cough, or washing hands often. Resuming normal activity is the preferred action. However, if both players at the two ends of an edge take action 1, then there is a chance of disease transmission. Accordingly, the constant $c_0$ captures a healthy player's sensitivity for avoiding a risky interaction. The constant $c_1$ captures a sick player's sensitivity to avoid transmitting the disease to one of its healthy neighbors. 
\end{example}

\begin{example}[Political polarization] Players want to differentiate their actions from those with opposing beliefs \cite{mcconnell2018economic}. The network represents the social interactions among players in opposing beliefs ($\ccalS_0$ and $\ccalS_1$). Action 1 represents a monetary choice or support for a cause that is individually desirable in the absence of partisanship. A player's tendency to take the preferred action (action 1) reduces as it has more neighbors that take action 1. That is, a player can opt-out from individual benefits or societal impact to express partisan preferences. Constants $c_0$ and $c_1$ capture the inclination of players in beliefs 0 and 1 to differentiate themselves from the players in the opposing belief, respectively.
\end{example}

\begin{example}[Hawk-Dove network game] Two competing species ($\ccalS_0$ and $\ccalS_1$) face-off in an ecological environment. At each interaction players decide to be hawkish ($a_i =1$) or dovish ($a_i = 0$). A hawk move gets the highest reward if its neighboring competitors play dove. If both interacting players play dove, they miss the opportunity to overcome their competitor. If both interacting players are hawkish, they challenge each other and face costs. The constants $c_0$ and $c_1$ represent the costs species 0 and 1 incur, respectively, when they act hawkish against a hawkish competitor. 
\end{example}

\begin{figure*}[t]
	\centering
		\includegraphics[scale=.95]{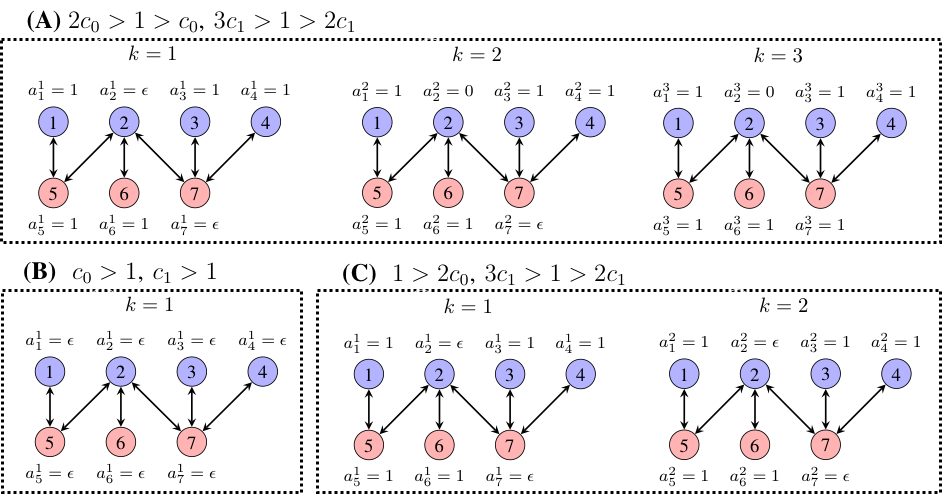}
\caption{Algorithm \ref{local_alg}'s convergence with respect to different payoff constants (A, B, C) on a 7 player network. Initially, all players are undecided $a_{i,0} = \epsilon$. {\bf(A)} At step $k=1$, players 1, 3, and 4 update their actions to 1 by \eqref{worst_case_good_response} given $1 > c_0$, players 5 and 6 also update their action to 1 by \eqref{worst_case_good_response} given $1> 2c_1$, and players 2 and 7 remain undecided. At step $k=2$, player 2 observes previous actions of its neighbors---$\{a_{5,1}, a_{6,1}\}$---and updates its action to 0 by \eqref{best_case_bad_response} given $2 c_0 > 1$. At step $k=3$, player 7 observes player 2's update---$a_{2,2}=0$---and updates its action to 1 by \eqref{worst_case_good_response} given $1 > 2 c_1$. All players are decided, and the action profile is the unique Nash equilibrium of the game. {\bf (B)} All players remain undecided for all times because no player can eliminate any action by \eqref{worst_case_good_response}. {\bf (C)} At step $k=1$, players 1, 3, 4, 5 and 6 update to action 1 similar to case (A). At step $k=2$, neither player 2 nor player 1 can eliminate any actions, and both remain undecided. A subset of players are decided while the others remain undecided. 
}
\label{fig:algorithm_example}
\end{figure*}
\section{Decentralized learning dynamics}\label{sec:learning}

Players repeatedly play the anti-coordination network game $\{\ccalN, \ccalA, \ccalG_B,\{u_i\}_{i\in \ccalN}\}$ taking actions $a_i^k\in \ccalA_i$ at each stage $k$. We assume each player knows its own type. Players take actions according to a decentralized algorithm details described in Algorithm \ref{local_alg}. 

Algorithm \ref{local_alg} starts with each player selecting an arbitrary action $\epsilon\in (0,1)$ not equal to zero or one. 
Player $i$ checks the worst and best possible outcomes in equations \eqref{worst_case_good_response}-\eqref{best_case_bad_response}, respectively. In \eqref{worst_case_good_response}, player $i$ checks whether it is preferable to select $a_i=1$ according to its utility \eqref{utility} even when its undecided neighbors ($\{j\in\ccalN_i:a_{j,k-1} = \epsilon\}$) end up taking action 1. That is, the ceiling operator makes a worst case scenario assumption (all undecided neighbors take action 1) and evaluates its utility from action 1. In \eqref{best_case_bad_response}, player $i$ checks whether it is preferable to select $a_i=0$ according to its utility \eqref{utility} when its undecided neighbors ($\{j\in\ccalN_i:a_{j,k-1} = \epsilon\}$) end up taking action 0. That is, the floor operator makes a best case scenario assumption (all undecided neighbors taking action 0) and evaluates $a_i=0$ according to \eqref{utility}. Note that the ceil and floor operators do not affect the decided neighbors of a player, i.e., neighboring players whose previous action are 0 or 1. If \eqref{worst_case_good_response}-\eqref{best_case_bad_response} do not hold at step $k$, player $i$ remains ``undecided'', i.e., $a_i^{k} = \epsilon$. 

\begin{algorithm}[Local learning algorithm]\label{local_alg}\hfill\\
{\bf Initialize:} $a_i^0 =\epsilon$ $\forall i \in \ccalN$ \\
{\bf for} $k=1,2,\dots$\\
Observe $a_{\ccalN_i}^{k-1}$  \\
\indent \indent 
\begin{align}
a_i^k &= 1 \quad{\textrm if }\quad 1 >  (c_0 (1-s_i)  +c_1 s_i) \sum_{j\in \ccalN_i} \left\lceil a_j^{k-1} \right\rceil
\label{worst_case_good_response}\\
a_i^k &= 0 \quad{\textrm if }\quad 1 <  (c_0 (1-s_i)  +c_1 s_i) \sum_{j\in \ccalN_i} \left\lfloor a_j^{k-1} \right\rfloor
\label{best_case_bad_response}\\
a_i^k&= \epsilon, \quad{\textrm otherwise}
\end{align}
\indent {\bf end}\\
{\bf end}
\end{algorithm}
The local algorithm takes as input an initial action profile $a^0$, and outputs an infinite sequence of action profiles. We denote the mapping of Algorithm \ref{local_alg} as
\begin{equation} \label{eq_action_profile}
	 \Phi(a^0) \equiv ( a^0, a^1, \ldots )
\end{equation}
where $a^t := (a_1^t,\ldots,a_n^t)$, We will use the notation 
\begin{equation} \label{eq_action_profile_k}
	\Phi_k(a^0) \equiv a^k
\end{equation}
to denote the resulting action profile after $k$ iterations of the local algorithm.

Figure \ref{fig:algorithm_example} shows the iterations of Algorithm \ref{local_alg} on a 7 player network for different payoff constants. We observe that depending on the payoff constants $c_0,c_1$, the algorithm might yield an action profile where all players are decided, all remain undecided, or some are decided and some remain undecided. In all cases, the algorithm converges in at most $k=3$ steps. 
Indeed, the algorithm will only require at most $n$ updates to converge where we recall $n$ as the number of players. Additional number of iterates is not needed because once a player decides on action $0$ or $1$, they do not revert. Further, if no player updates at a given step, then it implies there cannot be any updates in the future steps. This means some players can remain undecided. 
We present the convergence properties of Algorithm \ref{local_alg} in  the following section.

\section{Convergence of the local learning algorithm} \label{results_learning_dynamics}
%

\input{results_learning_dynamics.tex}

\section{Maximum anti-coordination problem}\label{sec:MPCAC}
We say the edge between players $(i,j) \in \ccalE_B$ is \emph{inactive} if $a_i  a_j = 0$, with $a_i,a_j \in \{0,1\}$. We have maximum anti-coordination when all edges are inactive, that is, $\sum_{(i,j)\in\ccalE_B} a_i  a_j = 0$ with $a_i,a_j \in \{0,1\}$. The  learning  dynamics  (Algorithm  \ref{local_alg}) does not  guarantee that the  resulting  action profile  satisfies  maximum  anti-coordination.  Hence,  to  ensure  maximum  anti-coordination,  it is  necessary  to  externally  control  players'  decisions.  To  this  end, we  formulate  the  minimum  player  control  for  maximum  anti-coordination  (MPCAC)  optimization  problem. 

Before we define the optimization problems, we define a controlled action profile trajectory in lieu of Algorithm \ref{local_alg}. A \emph{control profile} is an infinite sequence of subsets of players: $\ccalX = \{ \ccalX^0, \ccalX^1, \ldots \}$ with $\ccalX^t \subseteq \ccalN$ $\forall t \geq 0$. We say $x_i^t = 1$ if player $i \in \ccalX^t$, and $x_i^t = 0$ otherwise. Furthermore, we denote $\delta_i^t \in \{0,1\}$ as the \emph{forced action} of player $i$ at time $t\geq 0$ if $i\in\ccalX^t$. For convention, and without loss of generality, we say $\delta_i^t = 0$ if $i \notin \ccalX^t$. We write $\Delta \equiv (\delta^0,\delta^1,\ldots)$ for the sequence of forced action profiles. With the control profile $\ccalX$, the resulting action profile trajectory, with intial action profile $y^0$ is written
\begin{equation} \label{eq:controlled_dynamics}
	\Phi(y^0,\ccalX,\Delta) := (a^0,a^1,\ldots)
\end{equation}
where the $a^t$ obey the following dynamics for  $t = 0,1,\ldots$
\begin{equation}  \label{eq_dynamics}
	\begin{aligned}
		a_i^t &= (1-x_i^t)y_i^t + x_i^t \delta_i^t, \forall i \in \ccalN \\
		y^{t+1}&= \Phi_1(a^t) .
	\end{aligned}
\end{equation}
Note $\Phi_1(a^t)$ is the uncontrolled action at time $t+1$ by \eqref{eq_action_profile_k} given the controlled action profile $a^t$ at time $t$. We will refer to the pair $(\ccalX,\Delta)$ as a \emph{control policy}. In the first formulation of the maximum anti-coordination optimal control problem, we seek to achieve maximum anti-coordination using as few fixed control players as possible. It is formalized as follows.
\begin{definition}[Static MPCAC] \label{def_macac_1}
\begin{align}
\min_{\substack{\ccalX_s \subset \ccalN \\ \delta_i\in \{0,1\}, i\in \ccalX_s}} &J_s(\ccalX_s,\delta) := |\ccalX_s| \label{eq_static_objective}\\
&\text{s.t.} \\
&a_i^n+a_j^n \leq 1 \quad \forall (i,j) \in \ccalE_B\\
& a_i^n\in \{0,1\} \quad \forall i\in \ccalN  \\
& \ccalX = \{ \ccalX_s,\ccalX_s,\ldots \} \\
& \delta_i^t = \delta_i \in \{0,1\} \quad \forall i \in \ccalX_s, \forall t \geq 0 \\
& (a^0,a^1,\ldots) = \Phi(\vec{\epsilon},\ccalX,\Delta), \text{ where } \vec{\epsilon} = \epsilon 1_n
\end{align}

\end{definition}
The objective in \eqref{eq_static_objective} is to minimize the cardinality of the set $\ccalX_s$, that is, the number of players controlled. The first and second constraints together make sure of maximum anti-coordination at time $n$ when all players must be decided, i.e., they are equivalent to the constraint $\sum_{(i,j)\in\ccalE_B} a_i  a_j = 0$ with $a_i^n\in \{0,1\}$ for all $i\in\ccalN$. The third and fourth constraints define the controlled players and their actions for all times. The third constraint selects forced actions in the control set. The last constraint states the controlled learning dynamics \eqref{eq:controlled_dynamics} that govern players' decision-making.

The optimization problem \eqref{eq_static_objective} runs for $n$ time steps to allow for convergence of Algorithm \ref{local_alg} because if the algorithm dynamics are going to generate an anti-coordinating action pair, we would want to make use of it instead of incurring the cost for controlling a player. Selecting a player to the controlled set and changing its action via $\delta_i$ does not increase the convergence time of the algorithm, hence we stop the optimization after $n$ steps. Indeed, if we control a set of players $\ccalX_s$ at $t= 0$,  by Theorem \ref{local_algorithm_convergence} the algorithm converges in $n-|\ccalX_s|$ steps. Hence, the optimization horizon of $n$ steps is sufficient to make full use of learning dynamics.

The following result uses the convergence in finite time of the iterated elimination process to show that any static policy that satisfies the maximum anti-coordination constraint at time $n$ will continue to satisfy it for $t>n$. 

 \begin{lemma}\label{thm_no_cycle}
If $\Pi_s = (\ccalX_s,\Delta_s)$ is a static control policy such that $a_i^n + a_j^n\leq 1$ for all $(i,j) \in \ccalE_B$, then no player will change its decision, that is, $a_i^t + a_j^t\leq 1$ for all $(i,j) \in \ccalE_B$, $t>n$.
\end{lemma}
\begin{myproof}
A control profile $\ccalX_s$ with actions $\delta_i=\{0,1\}$ is feasible if all players decide by time $n$. Given a feasible static control policy $(\ccalX_s, \Delta)$, define the game $\Gamma'$ among players $\ccalN \setminus \ccalX_s$ where players connected to $\ccalX_s$ have a set of decided neighbors according to forced action profile $\Delta$.
The game $\Gamma'$ must be dominance solvable so that Algorithm \ref{local_alg} converges by time $n$. Thus if we continue to apply the forced actions $\Delta$, no player would change its decision after time $n$.
\end{myproof}
\\

In static MPCAC, we decide on players to control at the beginning and set their actions for the entire horizon. The objective only accounts for the number of players controlled  but not the number of times we control a player. In many situations it may be enough to control a player for a finite time to achieve maximum anti-coordination. For instance, in case (B) in Fig \ref{fig:algorithm_example}, if we set the actions of players 5-7 to 0 for one time step, the remaining players (1-4) will take action 1 by \eqref{worst_case_good_response}. If we stop controlling the players 5-7 in the next time step, they will continue to take action 0 by \eqref{best_case_bad_response}. Hence, the resultant action profile will achieve maximal anti-coordination. Next we formulate the dynamic MPCAC problem that allows for dynamic selection of players to control, and accounts for the number of times we control each player.

\begin{definition}[Dynamic MPCAC] \label{def_macac_3}
\begin{align}
\min_{\ccalX, \Delta} J_d(\ccalX,\Delta) &:= \frac{1}{n}\sum_{t=1}^{n} |\ccalX^t|+ \lim_{T' \rightarrow \infty} \frac{1}{T'}\sum_{t=n+ 1 }^{T'} |\ccalX^t| \label{eqn_objective}\\
&\text{s.t.} \\
&a_i^t+a_j^t \leq 1 \quad \forall (i,j) \in \ccalE_B, \; t\geq n\\
& a_i^t\in \{0,1\} \quad \forall i\in \ccalN, \; t\geq n  \\
& (a^0,a^1,\ldots) = \Phi(\vec{\epsilon},\ccalX,\Delta)
\end{align}
\end{definition}
The two terms in the penalty function in \eqref{eqn_objective} equally weight the control effort per player before convergence and after convergence to maximum anti-coordination. 
In dynamic MPCAC, we allow for the set of controlled players and their actions to change at each step. Lemma \ref{thm_no_cycle} does not necessarily apply in the dynamic setting. Hence, we explicitly require that the maximum anti-coordination is maintained for all times after $n$ in the first two constraints. The last constraint specifies the controlled learning dynamics \eqref{eq:controlled_dynamics}.

The following result shows that an optimal policy for dynamic MPCAC should at least be as good as an optimal policy for static MPCAC.
\begin{lemma} \label{thm_static_dynamic}
Let $\Pi_s = (\ccalX_s^*,\delta^*)$ and $\Pi_d = (\ccalX_*,\Delta_*)$ be optimal policies for static and dynamic MPCAC, respectively.  Then $J_d(\Pi_d) \leq 2J_s(\Pi_s) = 2|\ccalX_s^*|$. 
\end{lemma}
\begin{myproof}
Suppose there exists an optimal dynamic policy $\Pi_d$ such that $J_d(\Pi_d) >2 |\ccalX_s^*|$. Then we can implement $(\Pi_s)_{t\geq 0}$ to achieve a cost of $2 |\ccalX_s^*|$ in dynamic MPCAC, where the first and second terms in the objective \ref{eqn_objective} will be $|\ccalX_s^*|$. By Lemma \ref{thm_no_cycle}, $\Pi_s$ will satisfy the constraints in dynamic MPCAC which means it is also a feasible solution for the dynamic MPCAC. Hence, $\Pi_d$ cannot be optimal. 
\end{myproof}
\\
This result is expected when we observe that any static policy that is feasible for the static MPCAC is also feasible for the dynamic MPCAC by Lemma \ref{thm_no_cycle}. Hence, we can always use the optimal solution for static MPCAC to upper bound the penalty in the dynamic MPCAC. In fact, if the static MPCAC solution reaches an action profile that is an equilibrium of the game by time $n$, then the optimal solution is upper bounded by $|\ccalX_s^*|$. That is, we do not need to make any control efforts to remain at the maximum anti-coordination action profile because the action profile is also an equilibrium of the game.

The first constraint of dynamic MPCAC requires maximum anti-coordination after time $n$, whether or not control actions are used to keep it in equilibrium. This together with the penalization of control efforts for all times after $n$ gives preference to control policy solutions, when feasible, that leverage the controlled dynamics $\Phi$ in \eqref{eq:controlled_dynamics} in order to achieve maximum anti-coordination. The following result supports this intuition for dominance solvable games.

\begin{lemma} \label{thm_dominance_solvable}
If the game is dominance solvable, the optimal policy for dynamic MPCAC ($\Pi_d=(\ccalX_*, \Delta_*)$) is either $\ccalX^{t}_* = \emptyset$ for all $t=1,2,\dots$ or $\ccalX^{t}_* \neq \emptyset$ for $t\geq n$.
\end{lemma}
The above result proven in the Appendix shows that if Algorithm \ref{local_alg} converges to a single action profile, then either no control effort is required or the control effort after convergence ($t>n$) is non-zero. This result relies on whether the unique Nash equilibrium achieves maximum anti-coordination or not, which then corresponds to an empty control profile or a non-empty control profile for $t>n$, respectively. In general, if the game $\{\ccalN, \ccalA, \ccalG_B, \{u_i\}_{i\in\ccalN}\}$, not necessarily dominance solvable, has Nash equilibria that achieve maximum anti-coordination, we would expect control policies that induce convergence to such Nash equilibria over policies that control players after time $n$ in order to avoid the second term in the objective of dynamic MPCAC. We formalize this intuition in the following lemma (see appendix for the proof). 

\begin{lemma} \label{thm_anti_coordination_equilibria}
If there exists a set of Nash equilibria that achieves maximal anti-coordination, then the optimal solution to dynamic MPCAC will reach an action profile in this set.
\end{lemma}

Next we relate the dynamic MPCAC problem to a vertex covering problem on a bipartite network. We define the graph with potential dangerous links after Algorithm \ref{local_alg} converges at time $n$ as follows. Let $\ccalG^n = (\ccalN^n, \ccalE^n)$ be the graph with vertices composed of players that remain undecided or take action 1 after Algorithm \ref{local_alg} converges at step $n$, i.e., $\ccalN^n:=\{j\in\ccalN: a_j^n = \{\epsilon,1\}\}$, and with edges that connect players in $\ccalN^n$ in $\ccalE_B$, that is, $\ccalE^n = \{(i,j)\in \ccalE_B: i\in\ccalN^n, j\in\ccalN^n \}$. For example, in Fig. \ref{fig:algorithm_example} case (A), the network $\ccalG^n$ has the vertex set $\ccalN^n = \{1,3,4,5,6,7\}$. 

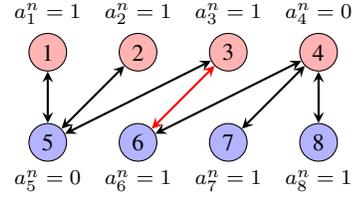
\begin{figure}[t]
	\centering
			\input{fig_3.tex}
\caption{Assume $c_0 =0.4$ and $c_1 = 0.4$. The game is dominance solvable and Algorithm \ref{local_alg} converges to the above action profile with only active link between 3 and 6. That is, $\ccalN^n = \{1,2,3,6,7,8\}$ and $\ccalE^n = \{(3,6)\}$. The solution to  \eqref{eq_objective_vertex_cover} if we disregard the last constraint is either $x_3^*=1$ or $x_6^* = 1$. These solutions violate the last constraint for players 5 or 4. If we fix $\rho_i=0$ for $i\in\{5,4\}$, then \eqref{eq_objective_vertex_cover} is infeasible. }
\label{fig_infeasible_example}
\end{figure}

A cardinality vertex cover for the graph $\ccalG^n = (\ccalN^n, \ccalE^n)$ looks for a minimum cardinality subset of vertices $\ccalX^n\subseteq \ccalN^n$ such that each edge has at least one endpoint incident at $\ccalX^n$ \cite{vazirani2013approximation}. Given the action profile at time $n$, $a^n$, we consider the following modified cardinality vertex cover problem with tuning parameter $\lambda\geq0$,
\begin{align}
\min_{\bbx, \bbrho\geq0}&  \sum_{i\in \ccalN^n} x_i + \lambda \sum_{i\in \ccalN\setminus \ccalN^n} |\rho_i| \label{eq_objective_vertex_cover}\\
\text{s.t.}\; &  x_i + x_j \geq 1 \quad \forall (i,j) \in \ccalE^n\\
& x_i \in \{0, 1\} \quad \forall i \in \ccalN^n\\
&c_i\sum_{j\in\ccalN_k} (\lceil a_j^n\rceil -x_j) \geq 1-\rho_i \; \forall k\notin \ccalN^n
\end{align}
where $c_i := c_0 (1-s_i) + c_1 s_i$ is player $i$'s payoff constant. If the second term in the objective and the last constraint are excluded, the problem formulation would be the minimum vertex covering problem in the bipartite network $\ccalG^n$.  If we set $\rho_i=0$ for players $i\in \ccalN \setminus \ccalN^n$, the last constraint makes sure that we select the players in the vertex covering such that the players in $\ccalN \setminus \ccalN^n$ who are decided on action 0 do not change their actions as a result of the control efforts. In general, the last constraint with $\rho_i=0$ can make the vertex covering problem infeasible---see Figure \ref{fig_infeasible_example} for an example. 

Let $\bbx^* = (x_1^*,\dots, x_n^*)$, and $\bbrho =(\rho_1^*, \dots, \rho_n^*)$ be an optimal solution to \eqref{eq_objective_vertex_cover}.  We construct a feasible control policy $\Pi_v=\{\ccalX,\Delta\}$ as follows. Let $\ccalX^t = \emptyset$ if $t<n$. Define the controlled player set $\ccalX^n$ and their actions $\Delta^n$ at time $n$ based on the optimal solution of the above problem as follows,
\begin{equation} \label{eq_controlled_set_n}
\ccalX^n = \{i\in\ccalN^n: x_i^* = 1\}\bigcup\{i\in\ccalN^n: x_i^* = 0, a_{i}^n = \epsilon\},
\end{equation}
and
\begin{equation} \label{eq_controlled_actions_n}
\delta_i^n = \begin{cases} 0 & \quad\forall i \in  \{i\in\ccalN^n: x_i^* = 1\}\\
1 & \quad\forall i \in  \{i\in\ccalN^n: x_i^* = 0, a_{i}^n = \epsilon\}.\end{cases}
\end{equation}

For $t=n+1$, we define the controlled player set $\ccalX^{n+1}$ and their actions $\Delta^{n+1}$   as follows,
\begin{equation} \label{eq_controlled_set}
\ccalX^{n+1} = \{i\in\ccalN^n: x_i^* = 1\}\bigcup\{i\notin\ccalN^n: \rho_i^*>0\},
\end{equation}
and
\begin{equation} \label{eq_controlled_actions}
\delta_i^{n+1} =  0 \quad\forall i \in \ccalX^{n+1}.
\end{equation}
For $t\geq n+1$, we let $\ccalX^t = \ccalX^{n+1}$ if the action profile $\{a_i^{n+1}, \delta_i^{n+1}\}_{i\in \ccalN}$ is not a Nash equilibrium of the game. Otherwise, $\ccalX^t =\emptyset$ for $t>n$. 
\begin{theorem} \label{thm_vertex_cover}
The dynamic control policy $\Pi_v =\{\ccalX,\Delta\}$ with control and action sets at time $n$ given by \eqref{eq_controlled_set_n}-\eqref{eq_controlled_actions_n} and for time $t\geq n+1$ given by \eqref{eq_controlled_set}-\eqref{eq_controlled_actions} is a feasible dynamic policy for the dynamic MPCAC problem in \eqref{eqn_objective}.
\end{theorem}
\begin{myproof}
Given \eqref{eq_controlled_set_n}-\eqref{eq_controlled_actions_n}, the controlled action profile at time $n$ is given by \eqref{eq_dynamics},
\begin{equation} \label{eq_controlled_profile_n}
a_i^{n} = \begin{cases} 1-x_i^* &\; \text{if } i \in \ccalN^n \\
0 &\; \text{o.w.} \end{cases}
\end{equation}
Given the definition of $\ccalG^n$ and $\Delta^n$, we have the controlled actions $a^n$ satisfy $a_i^n + a_j^n \leq 1$ for any $(i,j) \in \ccalE^n$, since we force at least one player in every link of $\ccalG^n$ to play action 0. This means that at time $n$ anti-coordination is achieved by $a^n$. Further, all players are decided satisfying the second constraint in \eqref{eqn_objective}.

Next, we show $\ccalX^{n+1}, \Delta^{n+1}$ maintains anti-coordination with the controlled action profile $a^{n+1}$.

Consider the following set partition of $\ccalN^n$,
\begin{align}
\ccalN^{n} &= \{i\in\ccalN: y_i^n = 1\}\\
&\cup \{i\in\ccalN: y_i^n = \epsilon, x_i^* = 1\}\\
& \cup \{i \in \ccalN: y_i^n = \epsilon,  x_i^* = 0\}\label{eq_N_n},
\end{align}
where $y^n$ is the uncontrolled action profile at time $n$. For the first two set of players in \eqref{eq_N_n}, we have $a_i^{n+1} = 1-x_i^*$ by \eqref{eq_controlled_set}. The last set of players were controlled to play action 1. At time $n$, they have no neighbors that take action 1 in \eqref{eq_controlled_actions_n}. So they continue to take action 1 using the update \eqref{worst_case_good_response} even when they are not controlled any longer. Hence, $y_i^{n+1} = 1-x_i^*$ for $\{i \in \ccalN: a_i^n = \epsilon,  x_i^* = 0\}$. This implies that $a_i^{n+1} = 1-x_i^*$ for all $i\in\ccalN^n$. 

Next, consider the following partition of the set of players not belonging to $\ccalN^n$,
\begin{align}
\ccalN\setminus&\ccalN^{n} = \{i\in\ccalN\setminus\ccalN^{n}: a_i^n = 0,c_i \sum_{j\in\ccalN_i} (\lceil a_j^n\rceil - x_j^*) < 1 \}\\
&\cup \{i\in\ccalN\setminus\ccalN^{n}: a_i^n = 0,c_i \sum_{j\in\ccalN_i} (\lceil a_j^n\rceil - x_j^*) > 1\} \label{eq_notin_N_n}
\end{align}
Note that all the players in the first set in \eqref{eq_notin_N_n} are controlled to play action 0 in \eqref{eq_controlled_set}, that is, $a_i^{n+1} = 0$ because $\{i\in\ccalN\setminus\ccalN^{n}: a_i^n = 0,c_i \sum_{j\in\ccalN_i} (\lceil a_j^n\rceil - x_j^*) < 1\} = \{i\notin\ccalN^n: \rho_i^*>0\}$ by the last constraint in \eqref{eq_objective_vertex_cover}. For the second set in \eqref{eq_notin_N_n}, all players continue to select action 0 by \eqref{best_case_bad_response}. This implies that $a_i^{n+1} = 0$ for $ i\notin \ccalN^n$. 

Combining the two arguments, we have $a_i^{n+1} = a_i^n$ where $a_i^n$ is given in \eqref{eq_controlled_profile_n}. Hence, the controlled action profile at time $n+1$ achieves anti-coordination. Suppose, $a^{n+1}$ is a Nash equilibrium action profile, then $a^{n+1} = \Phi_1(a^{n+1})$ and no further control effort is necessary. Otherwise, we have $y^{n+2} =\Phi_1(a^{n+1})=\Phi_1(a^{n})= y^{n+1}$ where $y^{n+1}$ and $y^{n+2}$ are the uncontrolled action profiles at time $n+1$ and $n+2$, respectively. 
Since, $\ccalX^t = \ccalX^{n+1}$, $\Delta^t=\Delta^{n+1}$ for $t\geq n+1$, we have that $a^{n+2} = a^{n+1}$. By induction, the control policy $\Pi_v$ satisfies anti-coordination. 
\end{myproof}

The proof relies on showing that a minimum vertex cover will eliminate all possible risky interactions after Algorithm \ref{local_alg} possibly eliminates a subset of them. The optimization formulation in \eqref{eq_objective_vertex_cover} makes sure that there are no changes to the players decided on action 0 by time $n$. The minimum vertex cover solution $\ccalX^n$ and forced actions $\delta_i^n=0$ for all $i \in \ccalX^n$ makes sure that the first two constraints in dynamic MPCAC are satisfied. The policy after time $n$ makes sure that all players continue to take the same actions for time $t>n$ hence satisfying the first two constraints for all $t>n$. 

Theorem \ref{thm_vertex_cover} shows that the MPCAC problems can be upper bounded by solving the minimum vertex cover on the reduced bipartite graph $\ccalG^n$. Further, if we set the tuning parameter $\lambda =0$, then the integer program in \eqref{eq_objective_vertex_cover} has an exact linear programming relaxation due to total unimodularity of bipartite networks \cite[Ch. 3]{nemhauser1988integer}. Hence, we can obtain a feasible policy efficiently by solving \eqref{eq_objective_vertex_cover} with penalty term $\lambda=0$. Then, including all players that have positive $\rho_i$ in the control set ($\{i\notin\ccalN^n: \rho_i^*>0\}$). The following corollary presents a scenario in which $\Pi_v$ optimal. 

\begin{corollary}\label{thm_optimal_vertex_cover}
The policy $\Pi_v$ defined in Theorem \ref{thm_vertex_cover} is an optimal policy if all players eliminate their actions in one time step in Algorithm \ref{local_alg}. 
\end{corollary}

The proof given in the appendix relies on showing that when all players eliminate their actions, the optimization problem in \eqref{eq_objective_vertex_cover} reduces to solving the minimum cardinality vertex cover for the entire network $\ccalG$. This case happens only when $c_i$ is smaller than inverse of  the maximum degree of the network ($c_i |\ccalN_i|<1$ for all $i\in\ccalN$). The control policy $\Pi_v$ is an upper bound in general because it does not make use of cascades, i.e., use the learning dynamics to make multiple hop links inactive, by controlling a subset of players. Instead, it waits for the algorithm to eliminate as many active links as possible, and then makes a two time-step control of players. As we show in numerical examples in Section \ref{sec:simulation}, the policy $\Pi_v$ tends to perform well when $c_0$ is small and $c_1$ is large, or when $c_0$ is large and $c_1$ is small. In the following section, we present optimal policies for benchmark networks that exemplify the optimal use of cascades to eliminate active links.

\section{Optimal solutions for benchmark networks}\label{sec_benchmark_optimal}
\input{Benchmark.tex}

\section{Suboptimal algorithms for general networks}\label{sec:greedy}

For general bipartite network topologies and for arbitrary network size, it becomes challenging to solve precisely for the MPCAC solutions (both static and dynamic). We devise an algorithm that selects at each iteration one player to control according to a greedy approach. The algorithm results in a subset of control agents $\hat\ccalX$ that ensures maximum anti-coordination, but may not be the optimal MPCAC solution. 

Given a network game $\{\ccalN, \ccalA, \ccalG_B, \{u_i\}_{i\in \ccalN}\}$, we define player $i$'s active neighbor set given action profile $a \in \ccalA$ as 
\begin{equation}
	\ccalN_i^*(a) =  \{ j \in \ccalN_i : a_j \in \{\epsilon,1\} \}.
\end{equation}
The \emph{active edge set} of the network in action profile $a$ is 
\begin{equation}
	\ccalE^*(a) :=  \{ (i,j) \in \ccalE_B : a_i, a_j \in \{\epsilon,1\} \}.
\end{equation}

At each iteration $k$, the \emph{greedy algorithm} selects the player $i^*_k$ that, upon holding its action fixed at $a_{i^*_k} = \delta_i \in \{0,1\}$,  results in the most number of active edges eliminated by the time the system dynamics converge.  We call the number of such links eliminated from choosing any player $i$ in the action profile $a$ the \emph{cascade potential} of player $i$,
\begin{equation}\label{eq:CP}
	\text{CP}_i(a,\delta_i) = |\ccalE^*(a)| - |\ccalE^*(\Phi_n(a,\{i\}_{t\geq0},\{\delta_i\}_{t\geq 0})) |.
\end{equation}
The process repeats, incrementally building up the player control set $\hat\ccalX$, until all active edges are eliminated from the network. The algorithm is detailed below.
\begin{algorithm}[Greedy algorithm]\label{suboptimal_alg}\hfill\\
{\bf Initialize:} $a_i^0 = \epsilon \  \forall i \in \ccalN$ \\
\indent\indent\indent\indent $\hat{\ccalX}(0) = \varnothing$ \\
\indent\indent\indent\indent $\hat{\delta} = 0_n$ \\
\indent\indent\indent\indent $k\gets 0$ \\
{\bf while} $\sum_{(i,j) \in \ccalE_B} a_i^k a_j^k \neq 0$\\
	\begin{enumerate}
		\item Allow dynamics to run until convergence \begin{equation} a^{k+1} \gets \Phi_n(a^k,\hat{\ccalX}(k),0_n) \end{equation} 
		\item Store convergence time \begin{equation} t_k \gets \min\{t \leq n : \Phi_t(a^k,\hat{\ccalX}(k),0_n) = \Phi_{t-1}(a^k,\hat{\ccalX}(k),0_n)\} \end{equation}
		\item Selection criterion \begin{equation} (i^*_k,\delta_{i^*_k}) \leftarrow \texttt{rand}\left( i \in \argmax_{\substack{ i : a_i^{k+1} \in \{\epsilon,1\} \\ \delta_i \in \{0,1\}}}  \text{CP}_i(a^{k+1},\delta_i)\right) \end{equation}
		\item Update control set \\ $\hat\ccalX(k+1) \gets \hat\ccalX(k) \cup i^*_k$ \\ $\hat\delta_{i^*_k} \gets \delta_{i^*_k}$
		\item $k \gets k+1$
	\end{enumerate}
\indent {\bf end}\\
$\hat\ccalX \gets \hat\ccalX(k)$ \\
{\bf end}
\end{algorithm}
If multiple players satisfy the maximum cascade potential criterion in step 3, ties are broken by random selection. The looping condition $\sum_{(i,j) \in \ccalE_B} a_i^k a_j^k \neq 0$ makes sure that we select players to control until no links remain active, that is, $a_i^k a_j^k = 0$ for all $(i, j)\in\ccalE_B$. 

We also consider a few variants of the above algorithm by replacing the selection criterion in step 3) by 
\begin{enumerate}[label=\alph*)]
	\item $i^*_k \gets \texttt{rand}(i \in \argmax_{i : a_i^k \in \{\epsilon,1\}} |\ccalN_i^*(a^k)|)$ \\ $\hat\delta_{i_k^*} \gets 0$  (max degree)
	\item $i^*_k \gets\texttt{rand}(\{i : (i,j) \in \ccalE^*(a^{k+1}) \text{ for some } j \})$ \\ $\hat\delta_{i_k^*} \gets 0$ (rand)
	\item $i^*_k \gets\texttt{rand}(i \in \argmax_{i,\delta_i} \text{CP}_i(a^{k+1},\delta_i) + \sum_{j=1}^n 1(a_j^{k+1} \neq a_j^k)$ (CP 2)
	\item Here, replace Algorithm \ref{suboptimal_alg} after {\bf{Initialize}} with 
	\begin{equation}
		\bar{a} \gets  \Phi_n(a^0) 
	\end{equation}
	\begin{equation}
		\hat\ccalX = \text{VC}(\ccalE^*(\bar{a}))
	\end{equation}
	where VC is the minimum vertex cover scheme detailed in \eqref{eq_controlled_set} of the resulting active network after the first convergence of the dynamics.
\end{enumerate}

In a), a player with the maximum number of active neighboring links is selected to play action 0. We denote this the ``max degree'' variant. In b), one player connected to the active network is selected uniformly at random with forced action 0. This algorithm called ``rand''. A variant of cascade potential, which we call ``CP2'' , is described in c), which selects the node that has the highest cascade potential in addition to inducing the most amount of players to change their action (see Remark \ref{alg_remark} for motivation). In d), we select the control set as the minimum vertex cover as specified in \eqref{eq_controlled_set} of Theorem \ref{thm_vertex_cover}. Note that in variants ``max degree", ``rand", and ``VC", the forced actions are always $\hat\delta_i = 0$, since it guarantees active edges to be eliminated (immediate neighbors). In the CP-based algorithms, $\hat\delta_i$ can either be 0 or 1, depending on which results in the most ``cascaded" active links broken.
\begin{proposition}
	For all  variants of Algorithm \ref{suboptimal_alg}, a resulting control set $(\hat\ccalX,\hat\delta)$  is a feasible static control policy of MPCAC. A feasible dynamic policy $(\ccalX,\Delta)$ is also produced by implementing control on the players $i_k$ in the order they were selected, and holding $a_{i_k} = 0$ for all times:
	\begin{equation}
		\begin{aligned}
			\ccalX^t &= \begin{cases} \hat\ccalX(k), \ \text{for } \sum _{m=0}^{k-1} t_m \leq t < \sum_{m=0}^{k} t_m \\ \hat\ccalX, \ \text{for } t \geq \sum_{m=0}^{\bar{k}} t_m \end{cases} \\
			\Delta^t &= \hat\delta, \  \forall t\geq 0
		\end{aligned}
	\end{equation}
	with $\sum _{m=0}^{-1} t_m \equiv 0$.
\end{proposition}
\begin{myproof}
	Consider the static policy $(\hat\ccalX,0_n)$. By the definition of Algorithm \ref{suboptimal_alg}, step 1 at the last iteration $k$ ensures maximum anti-coordination, where the control set used is $\hat\ccalX(k) = \hat\ccalX$. That is, $\Phi_n(a^k,\hat\ccalX(k),0_n)$ gives an action profile with maximum anti-coordination. It is also true that $\Phi_n(\vec{\epsilon},\hat\ccalX(k),0_n)$ gives the same action profile because the players in $\hat\ccalX$ were iteratively selected to eliminate all active links. If active links appeared as a result of a selection, these links are ensured to be eliminated in a subsequent iteration of the algorithm. Hence, $(\hat\ccalX,0_n)$ is a feasible static MPCAC solution.

The policy $(\ccalX^t,\Delta)$ is a feasible dynamic MPCAC solution because it simply mimics the iterations of Algorithm \ref{suboptimal_alg}.
\end{myproof}
\begin{remark}\label{alg_remark}
In practice, Algorithm \ref{suboptimal_alg} (CP variant, as written) will find the optimal solution to the star network of Proposition \ref{thm_star_staticMPCAC}. For the line network Proposition \ref{static_odd_line}, it finds the optimal solution in cases (b-e) always, and in cases (a,f) with some probability. In case (a), this is due to the possibility that the algorithm can select two players with different types along the line, which causes redundancies in control because the vertex cover $\ccalS_1$ is the optimal control choice by Corollary \ref{thm_optimal_vertex_cover}. In case (f), the algorithm could select a player one node from an endpoint (since it cascades the same number of links as any other player), when it could have selected the player two nodes from the endpoint, which covers more of the network and still causes the endpoint player to decide. This is the motivation behind introducing the CP2 variant - with this algorithm, it chooses the player two nodes from an endpoint with certainty because it causes the same number of cascaded links, but causes more players to change their decision.
\end{remark}

\begin{figure*}
	\centering
	\includegraphics[scale=.38]{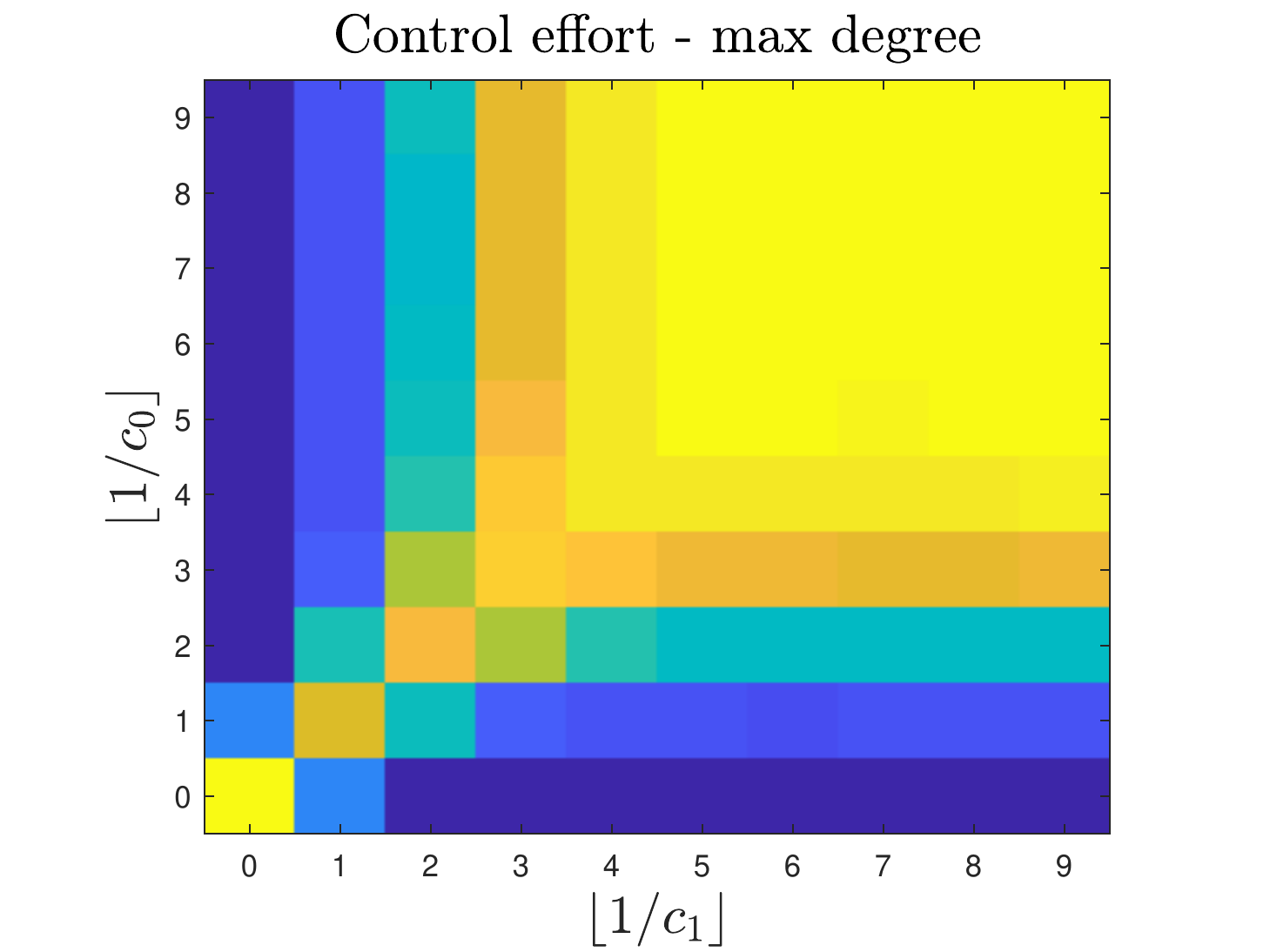}
	\includegraphics[scale=.38]{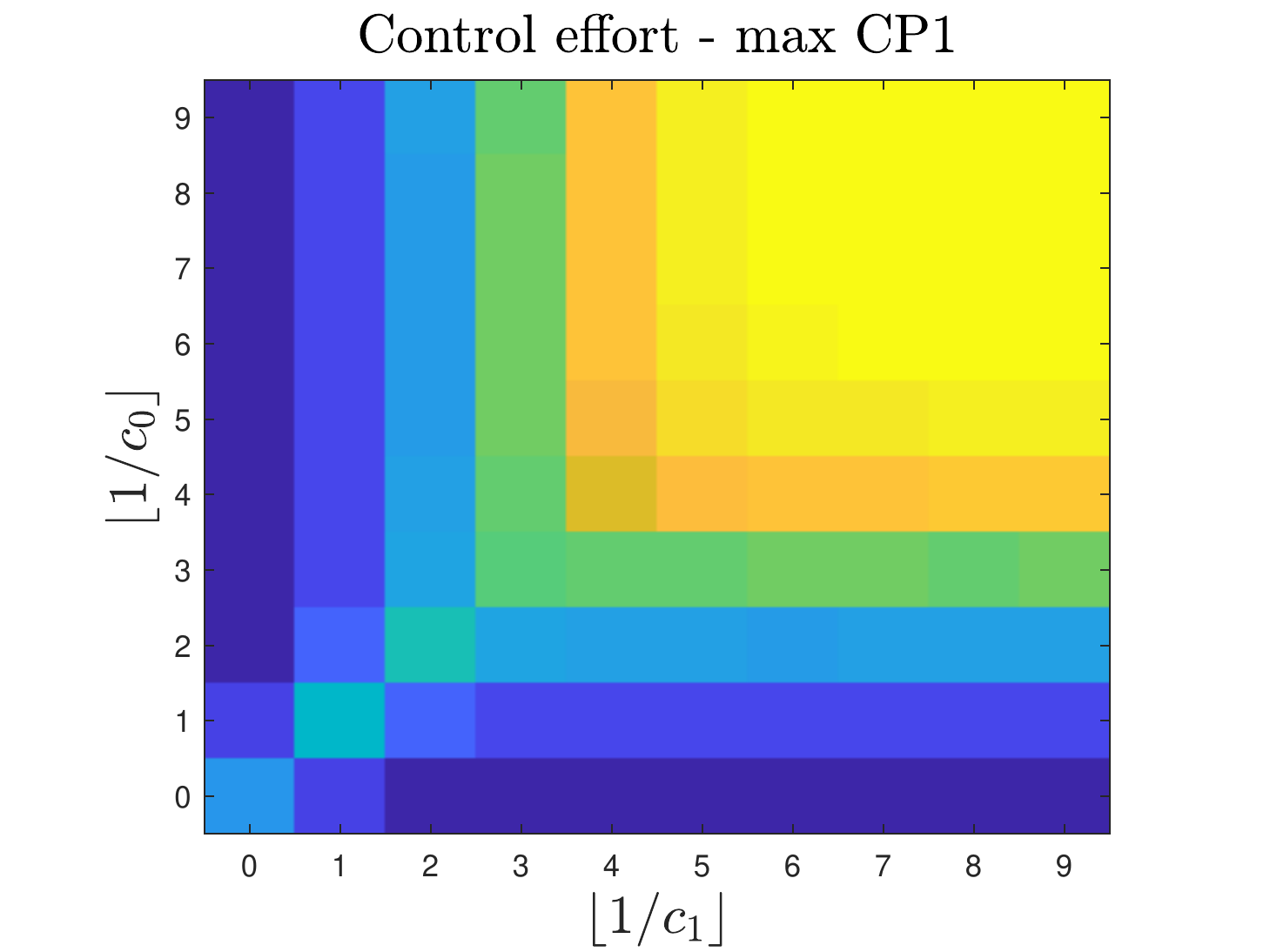}
	\includegraphics[scale=.38]{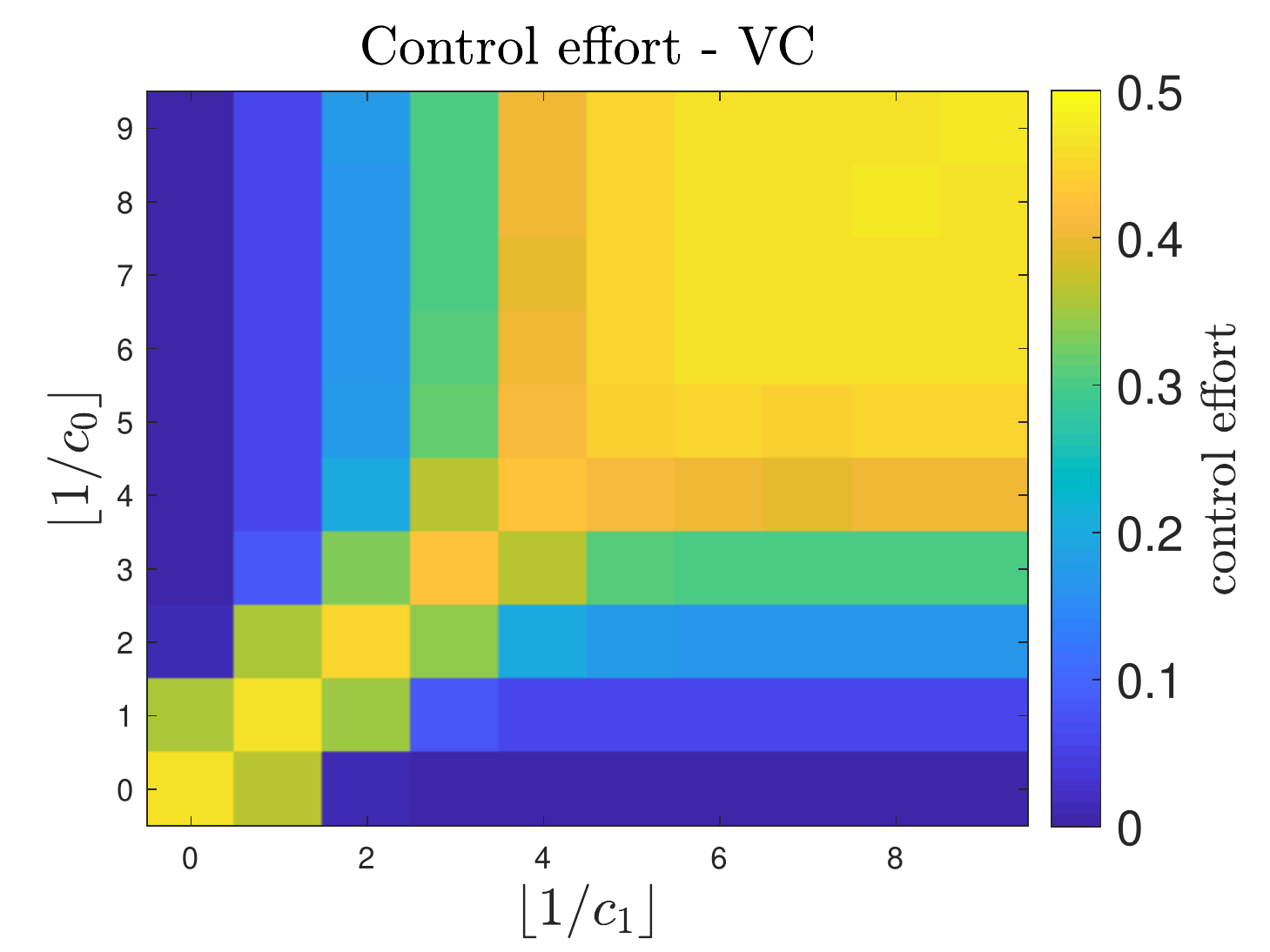}
	\caption{The control effort $|\hat\ccalX|/n$ used by Algorithm \ref{suboptimal_alg} on 20-node random bipartite networks, with $p_B = 0.3$ (expected degree of 3). Each payoff constant $c_0,c_1$ takes ten different values selected such that $m c_i > 1 > (m-1) c_i$ for $m = 1,\ldots,10$ and for $i = 0,1$. Each value in the grid results from averaging the resulting control effort from 1000 independent realizations of the network.}
	\label{fig:c0c1_maxdeg}
\end{figure*}
\begin{figure*}	
	\centering
	\includegraphics[scale=.35]{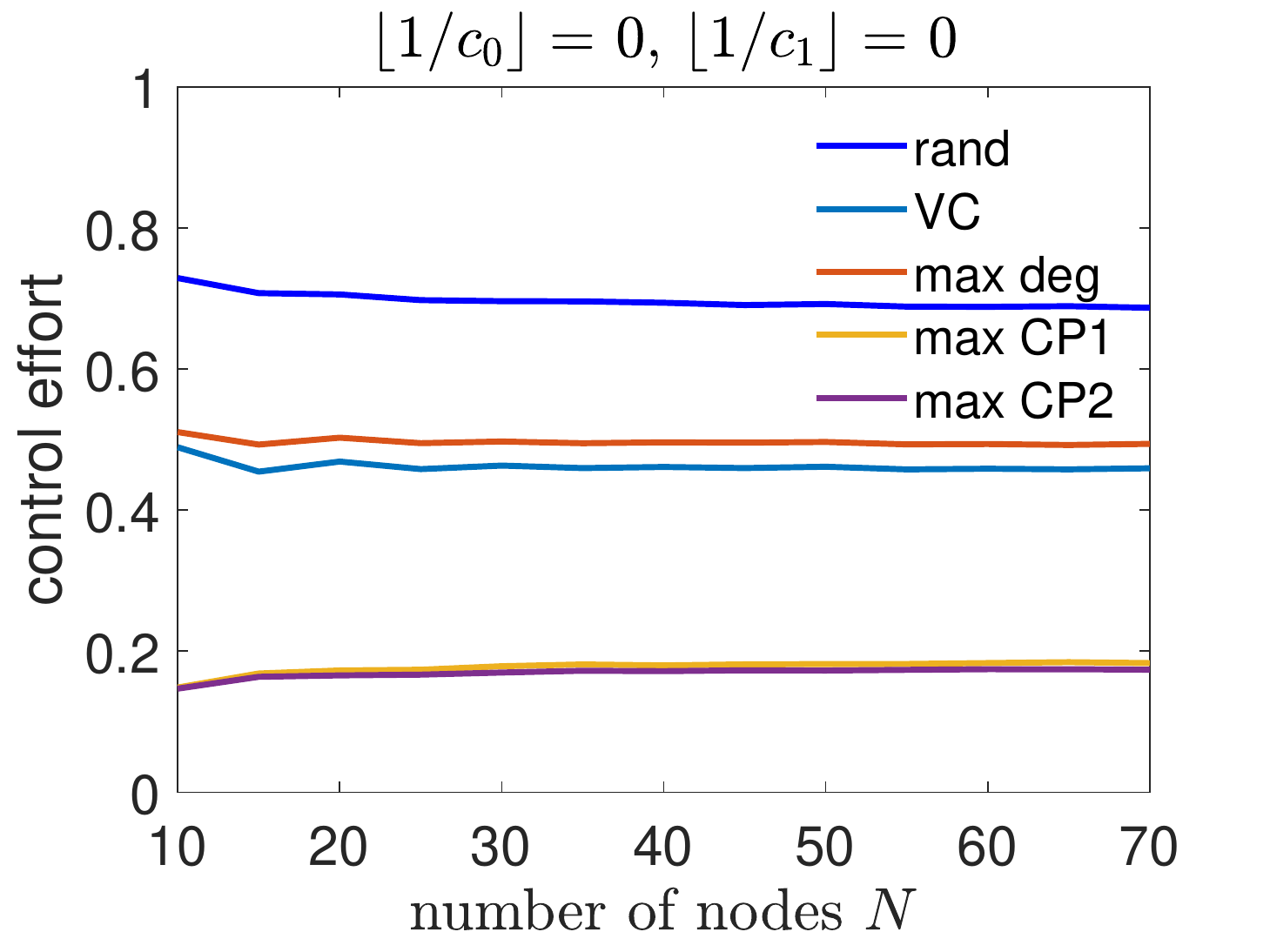}
	\includegraphics[scale=.35]{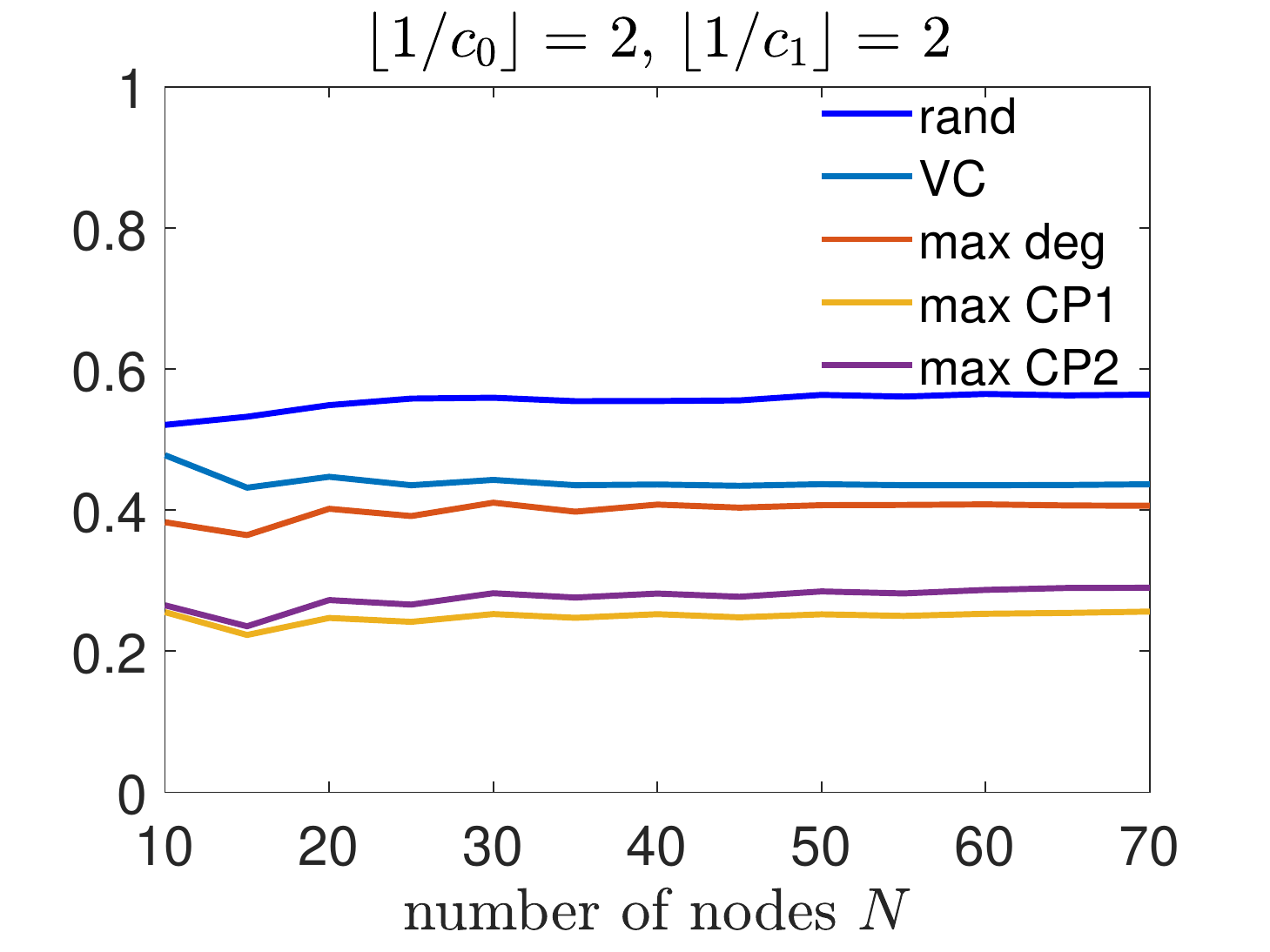}
	\includegraphics[scale=.35]{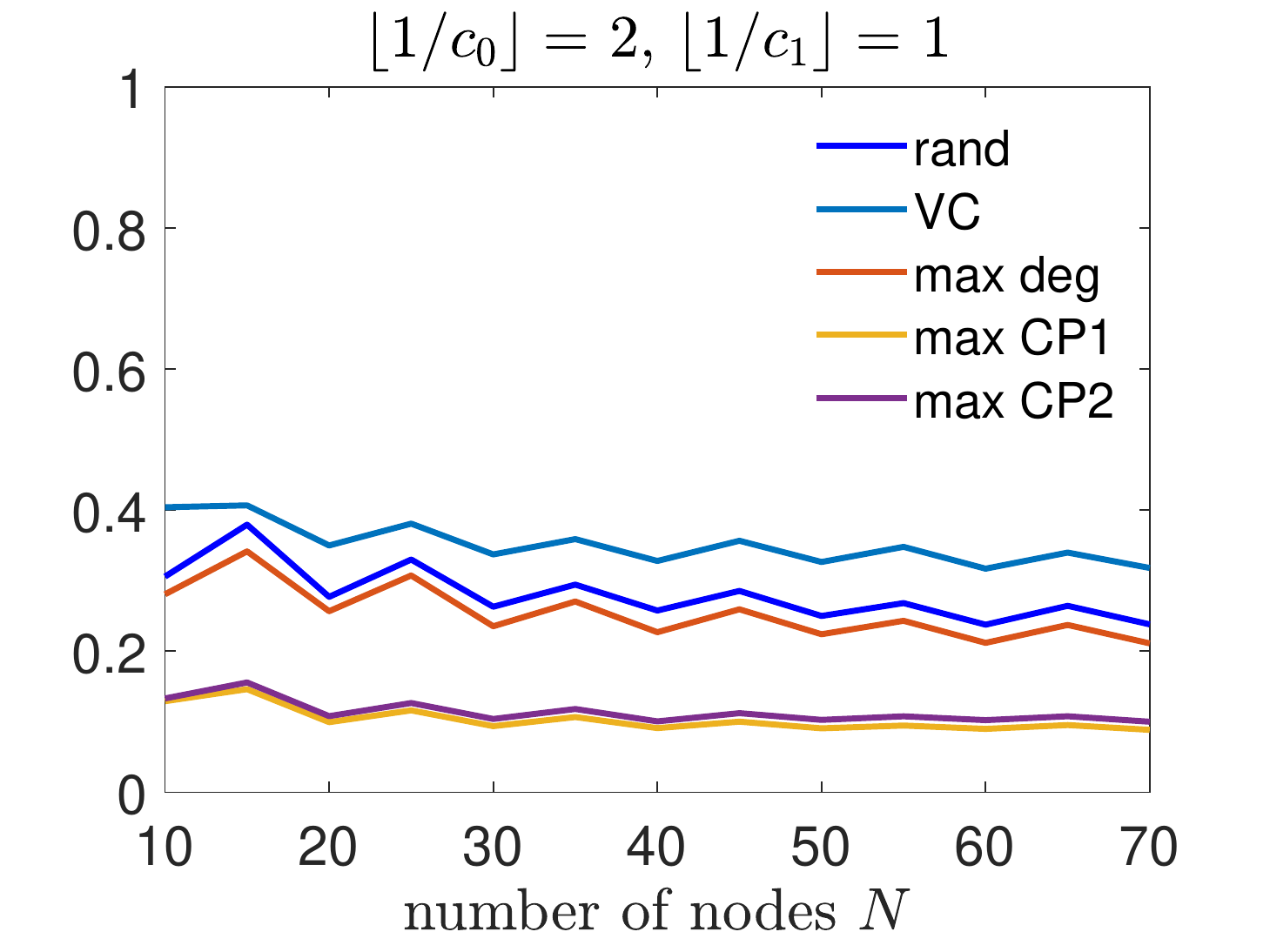}
	\caption{The fraction of nodes  $|\hat\ccalX|/n$ (control effort) to achieve maximum anti-coordination vs size of the network among the five variants of Algorithm \ref{suboptimal_alg}. The network is a bipartite network of size $N$, ranging from $N=10$ to $70$ with +5 increments. Each node is independently randomly chosen type 0 or 1  with equal probability, and the link $(i,j)$, $i \in S_0$ $j\in S_1$, is drawn with probability $6/N$, to achieve an expected degree of 3. The CP-based algorithms outperform the other variants.}
	\label{fig:suboptimal_comparison}
\end{figure*}

\section{Numerical Experiments} \label{sec:simulation}
We demonstrate the effectiveness of Algorithm \ref{suboptimal_alg} and its variants in numerical simulations over randomly generated bipartite networks. We generate a random bipartite network of size $n$ by assigning half of the players to $\ccalS_0$, and the other half $\ccalS_1$ (assuming $n$ is even). A link $(i,j)$ between $i \in \ccalS_0$ and $j \in \ccalS_1$ is present with independent probability $p_B \in (0,1)$.

In Figure \ref{fig:c0c1_maxdeg}, the control effort for three variants of Algorithm \ref{suboptimal_alg} is mapped over varying values of $c_0,c_1$. The quantity $\lfloor 1/c_0 \rfloor$ is the largest number of neighbors not playing action 0 a player in $S_0$ can have to still play action 1 as a dominant action. Similarly, $\lfloor 1/c_1 \rfloor$ is the largest number of neighbors not playing action 0 for an $S_1$ node. As both $c_0^{-1}$ and $c_1^{-1}$ increase, more control is necessary because more nodes will be playing dominant action 1. For $c_0^{-1}$ low and $c_1^{-1}$ high, no control effort is necessary to achieve maximum anti-coordination because $S_1$ nodes will play dominant action 1, and $S_0$ nodes in turn will choose dominant action 0. With both $c_0^{-1}, c_1^{-1}$ low, no nodes can initially decide. Hence, a large control effort is needed to cause the remaining nodes to play action 1.

In Figure \ref{fig:suboptimal_comparison}, we plot the performance of Algorithm \ref{suboptimal_alg} and its variants by measuring the control effort $|\hat\ccalX|/n$, the fraction of players selected. 
In the left panel, $\lfloor 1/c_0 \rfloor = \lfloor 1/c_1 \rfloor = 0$ - no players will decide without external control. Here, the CP-based algorithms perform significantly better than either variant because forcing $\delta_i = 1$ will cause all neighbors to play action 0. Max degree and VC have comparable performance here. In the center panel, $\lfloor 1/c_0 \rfloor = \lfloor 1/c_1 \rfloor = 2$ - a player has strictly dominant action 1 if two or less neighbors do not play action 0. The performance of the CP-based algorithms slightly worsen, while the variants slightly improve. In the right panel, $\lfloor 1/c_0 \rfloor = \lfloor 1/c_1 \rfloor = 1$. Here, the VC variant performs worse than ``rand", suggesting that the vertex covering scheme can  in some cases be inefficient as a control policy. Overall, the CP-based algorithms outperform all other variants due to their exploitation of ``cascades" by using the inherent dynamics $\Phi$ to eliminate active links.

\section{Conclusion}
\input{conclusion.tex}

\appendix
\input{appendix.tex}

\input{main.bbl}


\end{document}

%% file: my_sections.tex
\usepackage{needspace}



%% file: introduction.tex

Developing methods that achieve globally optimal behavior while conforming with the computational and informational limitations of the players is of interest given the ubiquity of noncooperative interactions that arise among actors in networked systems, e.g., epidemics \cite{nowzari2016analysis}, energy \cite{zhang2018dynamic,ye2017game}, security \cite{hota2018interdependent}, communication \cite{Bacci_et_al_2016} or autonomous systems \cite{marden2009cooperative}. Game theoretic learning algorithms are tractable decentralized models for non-cooperative decision-making in networked systems. These algorithms accounting for local information access \cite{Swenson_et_al_2014} and coupled action spaces \cite{yi2017distributed} guarantee convergence to individually rational behavior, i.e., a Nash equilibrium action, in certain classes of games, e.g., aggregative \cite{parise2015network,koshal2016distributed}, potential \cite{swenson2018distributed}, convex \cite{scutari2010convex}. However, a Nash equilibrium, while being optimal from the perspective of selfish individuals, can be inefficient and undesired at the system level. A canonical example of this is the tragedy of the commons which describes the phenomenon of selfish learning behavior leading to the worst possible outcome for the entire population \cite{hardin1968tragedy}. Given the possibility of emergence of undesired outcomes, there is a need to develop incentive mechanisms in order to achieve system-wide desired outcomes. 

The major challenge in attaining globally desired outcomes in networked systems is that individuals are selfish, heterogeneous, and their actions are coupled while the centralized incentive resources are costly. In this paper, we formulate this challenge as the control of decentralized learning dynamics. That is, players selfishly follow some game-theoretic learning dynamics while a centralized authority aims to direct the emergent behavior toward a desired outcome. The two main issues we address with this formulation are the selection of which players to control, and what control policy to implement given the selected players. 

This challenge is addressed in the literature by characterizing the inefficiency of Nash equilibria \cite{marden2014generalized}, by designing payoffs prior to start of the game to induce efficient Nash equilibria \cite{Li_Marden}, or by developing control mechanisms \cite{balcan2009improved,balcan2013circumventing,grammatico2017dynamic,riehl2018incentive,fele2018coalitional,aral2018social,Borowski_2018}. Our approach falls into the last category of controlling players to guide the learningdynamics towards desirable outcomes. In this category, \cite{balcan2009improved,balcan2013circumventing} show a public advertising scheme improves the efficiency of the emergent outcome for players that act according to best-response dynamics and occasionally listen to the advertised behavior.  An alternative model designs dynamic control incentives that affect every players' payoffs to which players best respond \cite{grammatico2017dynamic}. 
Minimum cost uniform and targeted reward policies that induce complete coordination among players that act according to best-response dynamics in a network coordination game are developed in \cite{riehl2018incentive}. When the goal is to minimize efficiency, \cite{Borowski_2018} studies malicious attacks that strategically perturb player learning dynamics in a network coordination game. In a combined effort to select which players to control and also how to control them, we depart from these studies by characterizing control schemes when the underlying interaction between players is an anti-coordination game.

In this paper, we focus on controlling a subset of the players in a network anti-coordination game with the aim to promote maximum anti-coordination. Players belong to one of two possible types. We assume there is a preferred selfish action for each player in the absence of any neighboring players in the network. That is, a player's payoff decreases as more of its neighbors belonging to the opposite type take the preferred action (Section \ref{sec:model}). Such payoff dependencies can be used to model individual behavior during the spread of an epidemic in a population where the individual types are healthy and sick, the actions represent the level of precautionary measures taken, and payoffs capture the risk of disease transmission to healthy from sick individuals with the preferred action being not taking any measures \cite{eksin2017disease}. Other examples include modeling individual opinions in a  politically polarized environment where players would like to differentiate their actions from players in opposing views \cite{mcconnell2018economic}, or modeling two competing species in an environment \cite{hauert2004spatial}. In these games, two neighboring players anti-coordinate when at least one of the players do not take the selfish action, e.g., one player in each link takes a precautionary measure during an epidemic. Maximum anti-coordination is achieved when there does not exist any link with failed anti-coordination. 

We assume players follow learning dynamics based on a decentralized process of iterated elimination of strictly dominated actions in the absence of any control \cite{Fudenberg_Levine_1998} (Section \ref{sec:learning}). In the learning process, players eliminate all actions that cannot be in a rational action profile in finite time (Theorem \ref{local_algorithm_convergence}). This implies convergence to the unique Nash equilibrium in dominance solvable games. 

We formulate the minimum player control for anti-coordination problem (MPCAC) as a mixed integer program where the decision variables include which players to control, and how to control them (Section \ref{sec:MPCAC}). We consider static and dynamic control policies. In static MPCAC, a fixed subset of players' decisions are controlled for the entire learning horizon. In dynamic MPCAC, the control policy can temporally influence player decisions. We find a feasible policy that upper bounds the optimal dynamic MPCAC control policy by solving a minimum cardinality vertex covering problem on a reduced bipartite network (Theorem \ref{thm_vertex_cover}). This feasible policy can be obtained by solving a linear program, and hence is computationally feasible to compute. 

In general, the vertex cover-based control policy is sub-optimal if there exists a way to cause cascades of anti-coordination among multi-hop neighbors via the learning dynamics by only controlling a few players. We  present optimal policies on benchmark networks of arbitrary size for every possible payoff constant values  (Section \ref{sec_benchmark_optimal}). Some of the optimal policies exhibit use of cascades to achieve anti-coordination which exemplify the sub-optimality of the vertex cover-based control policy. Based on the cascade inducing policies, we also propose a greedy algorithm that finds a feasible solution by sequentially selecting the player to control with the highest potential for inducing a cascade of anti-coordination (Section \ref{sec:greedy}). We compare the greedy algorithm with the vertex cover-based control policy, among several other variants, in numerical experiments on random bipartite networks (Section \ref{sec:simulation}).

%% file: fig_1.tex

\usetikzlibrary{matrix,arrows,decorations.pathmorphing}
\usetikzlibrary{arrows,automata}

{\small
\tikzstyle{sets} = [ellipse, draw=black, inner sep=0pt, minimum size=0.5cm]
\tikzstyle{terminal} = [circle, draw=black, inner sep=0pt, minimum size=0.5cm]
\tikzstyle{theta} = [circle, draw=white, inner sep=0pt, minimum size=0.5cm]
\tikzstyle{arrow} = [stealth-stealth, thick]
\def\boundellipse {(0,0) ellipse (10,5)}

\begin{tikzpicture}[x=0.4cm, y=0.3cm]
\draw
       ++( 0:0) node (terminal 1)  [terminal, fill=red!30] {1} 
	++( 0:3) node (terminal 2)  [terminal, fill=red!30] {2}
        ++( 0:3) node (terminal 3)  [terminal, fill=red!30] {3};
        
\draw
       ++( 270:4) node (terminal 4)  [terminal, fill=blue!30] {4} 
	++( 0:3) node (terminal 5)  [terminal, fill=blue!30] {5}
        ++( 0:3) node (terminal 6)  [terminal, fill=blue!30] {6}
         ++( 0:3) node (terminal 7)  [terminal, fill=blue!30] {7};

\draw 
	++(0:-3) node (theta 1) [theta, label = 0:\normalsize{$\ccalS_1$}]{}
	++(270:4) node (theta 2) [theta, label = 0:\normalsize{$\ccalS_0$}]{};

\draw (0,-4) [dashed,rotate=90] ellipse (.5cm and 1.7cm);
\draw (-3,-6) [dashed,rotate=90] ellipse (.5cm and 2.3cm);

%
\draw [arrow] (terminal 4) to [bend right = 0] (terminal 1);
\draw [arrow] (terminal 5) to [bend left = 0] (terminal 1);
\draw [arrow] (terminal 4) to [bend left = 0] (terminal 2);
\draw [arrow] (terminal 6) to [bend right = 0] (terminal 2);
\draw [arrow] (terminal 6) to [bend right = 0] (terminal 3);
\draw [arrow] (terminal 7) to [bend right = 0] (terminal 3);


\end{tikzpicture}
}

%% file: results_learning_dynamics.tex

\subsection{Game theoretic preliminaries}
We define the rational action by the Nash equilibrium solution concept. A Nash equilibrium (NE) action profile $a^{NE} \in [0,1]^n$ is such that no individual has a preferable deviation from its action, that is,
\begin{align} \label{Nash_definition}
u_i(a_i^{NE}, a_{\ccalN_i}^{\text{\text{NE}}}) \geq  u_i(a_i, a_{\ccalN_i}^{\text{NE}}) \quad a_i \in [0,1], i \in \ccalN
\end{align}
where $a_{\ccalN_i}^{\text{NE}} := \{a_{j}^{\text{NE}}: j\in \ccalN_i \}$.
In other words, the individuals respond to the NE actions of other individuals to maximize their payoffs. For a given neighbor action profile $a_{\ccalN_i}$ we have the best response of individual $i$ as follows,
\begin{align}
BR_i(a_{\ccalN_i}) &:= \argmax_{a_i \in [0,1]} u_i(a_i, a_{\ccalN_i}) \label{best_response} \\
& = \bbone\left({1} >  (c_0 (1-s_i)+c_1 s_i) \sum_{j\in \ccalN_i} a_j \right) \label{best_response_specific}
\end{align}
where $\bbone(\cdot)$ is an indicator function. Since the payoffs are linear in self-actions, the actions that maximize the payoffs are in the extremes---$a_i =1$ or $a_i = 0$---depending on the types and actions of their neighbors. 
We can equivalently represent the NE definition in \eqref{Nash_definition} by using the best response definition,  
\begin{align} \label{NE_best_response}
a_i^{NE} = BR_i(a_{\ccalN_i}^{\text{NE}}) \quad \forall i\in \ccalN.
\end{align}

The notion of strictly dominated is defined as follows.

\begin{definition}[Strictly dominated action] \label{strictly_dominated_definition}
An action $a_i \in [0,1]$ is strictly dominated if and only if there exists an action $a_i' \in [0,1]$ such that 
\begin{equation}\label{eq:strictly_dominated_definition}
u_i(a_i', a_{\ccalN_i}) >  u_i(a_i, a_{\ccalN_i}) \quad \forall a_{\ccalN_i}\in \ccalA_{\ccalN_i}
\end{equation}
\end{definition}
If an action $a_i$ is strictly dominated then there exists a more preferable action $a_i'$ for any circumstance. It is clear that if an action is strictly dominated then it cannot be a rational action from \eqref{NE_best_response}. 
%

In a game we can iteratively remove the strictly dominated actions, this process is called the iterated elimination of strictly dominated strategies and is defined below.
\begin{definition}[Iterated elimination] \label{def:elimination_strict}
Set the initial set of actions $A_i^0 = [0,1]$ for all $i$, and for any $k\in \naturals$ let 
\begin{equation}
	\begin{aligned}
	A_i^k = \{a_i \in &A_i^{k-1} \given \\
	&a_i \text{ is not strictly dominated by any } a_{\ccalN_i} \in A_{\ccalN_i}^{k-1}\}
	\end{aligned}
\end{equation}
\end{definition}
We denote the set of player $i$'s actions that survive the iterated elimination by $A_i^\infty := \bigcap_{k =0}^\infty A_i^k$. When $A_i^\infty$ has a single element, we say $A_i^\infty$ is a singleton. If $A_i^\infty$ is a singleton for all $i\in\ccalN$, then the game is dominance solvable, and has a unique Nash equilibrium given by the action profile that survives the iterated elimination process. 

\subsection{Convergence}

The following theorem states that Algorithm \ref{local_alg} eliminates all strictly dominated actions in finite number of steps.

\begin{theorem} \label{local_algorithm_convergence}
Algorithm \ref{local_alg} converges in at most $n$ iterations, that is, no player changes its action after the $n$th update. At the end of $n$ iterations, if all players are decided, i.e., $a_{i}^n = \{0,1\}$ for all $i\in\ccalN$, then all the other actions $\ccalA_i \setminus a_{i}^n$ are strictly dominated and the resultant action profile is a Nash equilibrium. Otherwise, if a player $i \in \ccalN$ is undecided, i.e., $a_{i}^n = \epsilon$, then there does not exist $a_i \in \ccalA_i$ that can be strictly dominated. 
\end{theorem}

The proof of the theorem is given in the appendix. It relies on showing that Algorithm \ref{local_algorithm_convergence} is a decentralized version of the iterated elimination of strictly dominated actions as given by Definition \ref{def:elimination_strict}. The intuition for $n$ step convergence is that at each step at least one player needs to eliminate its action using \eqref{worst_case_good_response} or \eqref{best_case_bad_response}. If no player updates at a time step, the players stop updating because no new eliminations are triggered from then on. Therefore, there could  at most be $n$ iterations to rule out $n$ players one player at a time.

%% file: fig_3.tex

\usetikzlibrary{matrix,arrows,decorations.pathmorphing}
\usetikzlibrary{arrows,automata}

{\small
\tikzstyle{terminal} = [circle, draw=black, inner sep=0pt, minimum size=0.5cm]
\tikzstyle{theta} = [circle, draw=white, inner sep=0pt, minimum size=0.5cm]
\tikzstyle{arrow} = [stealth-stealth, thick]

\begin{tikzpicture}[x=0.4cm, y=0.3cm]

\draw
       ++( 0:0) node (terminal 1)  [terminal, fill=red!30,label = 90:\footnotesize{{$a_{1}^n=1$}}] {1} 
       ++( 0:3) node (terminal 2)  [terminal, fill=red!30,label= 90:\footnotesize{{$a_{2}^n=1$}}] {2} 
       ++( 0:3) node (terminal 3)  [terminal, fill=red!30,label= 90:\footnotesize{{$a_{3}^n=1$}}] {3} 
       ++( 0:3) node (terminal 4)  [terminal, fill=red!30,label= 90:\footnotesize{{$a_{4}^n=0$}}] {4} ;

\draw
       ++( 270:4) node (terminal 5)  [terminal, fill=blue!30, label= 270:\footnotesize{{$a_{5}^n=0$}}] {5} 
       	++( 0:3) node (terminal 6)  [terminal, fill=blue!30, label= 270:\footnotesize{{$a_{6}^n=1$}}] {6}
	++( 0:3) node (terminal 7)  [terminal, fill=blue!30, label= 270:\footnotesize{{$a_{7}^n=1$}}] {7}
        ++( 0:3) node (terminal 8)  [terminal, fill=blue!30, label= 270:\footnotesize{{$a_{8}^n=1$}}] {8};
                
\draw [arrow,red] (terminal 3) to [bend right = 0] (terminal 6);
\draw [arrow] (terminal 3) to [bend right = 0] (terminal 5);
\draw [arrow] (terminal 5) to [bend left = 0] (terminal 1);
\draw [arrow] (terminal 5) to [bend left = 0] (terminal 2);
\draw [arrow] (terminal 6) to [bend right = 0] (terminal 4);
\draw [arrow] (terminal 7) to [bend right = 0] (terminal 4);
\draw [arrow] (terminal 8) to [bend right = 0] (terminal 4);
\end{tikzpicture}
}

%% file: Benchmark.tex

We consider star, line and ring networks, and provide optimal solutions for both static and dynamic MPCAC problems. 

\begin{figure*}[t]
	\centering
	\begin{tabular}{ccc}
			\input{star.tex} &
			\input{line.tex}&
			\input{ring.tex}
	\end{tabular}
\caption{Star, line, and ring benchmark networks for 4 player networks. }
\label{fig:benchmark_network}
\end{figure*}

\subsection{Star network}

Consider a star network of $n$ players with player 1 in the center. Without loss of generality we assume center node is type 1, i.e., $s_1 = 1$ and the rest are type 0, i.e., $j \in \ccalS_0$ for $j \in \ccalN\setminus \{1\}$---see Figure \ref{fig:benchmark_network}. 

\begin{proposition} \label{thm_star_staticMPCAC}
Depending on the utility constants $c_0$ and $c_1$, the optimal solutions to static MPCAC follows. 
\begin{itemize}
\item[a)] If $1>c_0$ and $1>c_1 (n-1)$, then $\ccalX = \{1\}$ with $\delta_1 = 0$.
\item[b)] If $1<c_0$ and $1>c_1 (n-1)$, or $1>c_0$ and $1<c_1 (n-1)$, then $\ccalX = \emptyset$.
\item[(c)] If $1<c_0$ and $1<c_1 (n-1)$, then $\ccalX = \{1\}$ with either $\delta_1 = 0$ or $1$.
\end{itemize}
\end{proposition}
\begin{myproof}
In a star network, if $1>c_1 (n-1)$ or $1>c_0$, then the center or the fringe players have all actions strictly dominated by action 1 via \eqref{worst_case_good_response}. If $1<c_0$, then a fringe player eliminates all actions except 0 via \eqref{best_case_bad_response} when the center node takes action 1. If $1<c_1 (n-1)$, then the center player eliminates all actions except 0 via \eqref{best_case_bad_response} when the fringe nodes take action 1. Based on this, in case (a), the algorithm converges to an action profile where all players take action 1. In order to achieve maximum anti-coordination, we select center player to play action 0. In case (b) the algorithm converges in 2 steps to a maximum anti-coordination action profile, i.e., $\{a_1 = 1, a_{-1} = 0\}$ or $\{a_1 = 0, a_{-1} = 1\}$. In case (c), there is no action that is strictly dominated if we do not control a player. Hence, we control center player to take action 0, i.e., $\ccalX=\{1\}$. Then, all fringe players eliminate all actions except action 1. 
\end{myproof}

Above result addresses the three cases that can arise in a star network: (a) all players make decisions using the algorithm but the resultant action profile does not achieve maximum anti-coordination; (b) all players make decisions using the algorithm and the resultant action profile achieves maximum anti-coordination; (c)  all players remain undecided using the algorithm. In case (a), we have to cause anti-coordination by controlling the center player. In case (b), we do not have to control any player. In case (c), we control the center player to trigger a decision for the fringe players. 

For the dynamic MPCAC, we need to specify the set of nodes controlled at each time step denoted by $\ccalX^t$. We provide optimal policies for dynamic MPCAC problem on a star network. 

\begin{proposition} \label{thm_star_dynamicMPCAC}
Depending on the utility constants $c_0$ and $c_1$, the optimal solutions to dynamic MPCAC follows. 
\begin{itemize}
\item[(a)] If $1>c_0$ and $1>c_1 (n-1)$, then
\begin{equation}
	\begin{cases}\ccalX^t =\varnothing, \ &\text{if } t < n \\ \ccalX^t =\{1\}, \delta_1^t = 0 &\text{if } t\geq n  \end{cases}
\end{equation}
The resulting optimal cost is $C^* = 1$.
\item[(b)] If $1<c_0$ and $1>c_1 (n-1)$, or $1>c_0$ and $1<c_1 (n-1)$, then $\ccalX^t = \varnothing$ for all $t$.
The resulting optimal cost is $C^* = 0$.
\item[(c)] If $1<c_0$ and $1<c_1 (n-1)$, then $\ccalX^1 = \{1\}$.
\begin{equation}
	\begin{cases}\ccalX^t =\{1\}, \delta_1^t = 0 \ &\text{if } t =0,1 \\ \ccalX^t =\varnothing &\text{if } t > 1  \end{cases}
\end{equation}
The resulting optimal cost is $C^* = 2/n$.
\end{itemize}
\end{proposition}
\begin{myproof}
In cases (a) and (b), the game is dominance solvable. Further, the algorithm converges in $t\leq 2$ steps. Lemma \ref{thm_dominance_solvable} together with Lemma \ref{thm_vertex_cover} gives the optimal policies for the first two cases. 

In case (c), the game is not dominance solvable and there exists two Nash equilibria that achieve maximum anti-coordination: $(a_1=0,\{a_i=1\}_{i\neq 1})$ and $(a_1 = 1, \{a_i=0\}_{i\neq 1})$. By Lemma \ref{thm_anti_coordination_equilibria}, the optimal policy will induce the dynamics to converge to one of these configurations. Either of these equilibria can be achieved through a dynamic policy that targets only the center node for a single time step, which gives the optimal cost $1/n$. For instance, we set the center player's action to 0 at time $t=0$ so that fringe players select $a_i=1$ at $t=1$.  We hold this control at $t=1$ as well, so that the center node will consolidate its unforced action to 0 by \eqref{best_case_bad_response}. We lift the control at $t=2$ and for all times thereafter, in which the players are in equilibrium and maximum anti-coordination is achieved. This policy is optimal because there cannot be a policy that achieves a smaller objective value than $2/ n$ and achieve maximum anti-coordination.
\end{myproof}

The optimal action profile and its proof are similar to the static case. There are two significant differences between the dynamic and static problem. If the algorithm converges to an action profile in which some neighboring pairs of players end up taking action 1, e.g., case (a), then we have to anti-coordinate the corresponding links by controlling one of the players at the end of the time horizon $n$. If the algorithm is going to converge to an action profile where we will have undecided neighboring players (case (c)), we can make these players anti-coordinate by initially controlling one of the neighbors temporarily for two time steps. In case (a), we incur a cost of 1 because if we lift our control of the center player, it will always revert back to $a_1=1$. In contrast, in case (c), we only need to control the player $1$ for two steps, because it will not revert to $a_1 = 1$ by the time its neighbors are taking action 1. 

\subsection{Line network}

 We consider a line network in which the neighborhood of player $i\in\ccalN \setminus \{1,n\}$ is given by $\ccalN_i=\{i-1,i+1\}$. The players at the endpoints have neighbor sets $\ccalN_1 = \{2\}, \ccalN_n = \{n-1\}$.  The type configuration alternates between types 0 and 1: $i \in \ccalS_0$ for $i$ odd, and $i\in\ccalS_1$ for $i$ even. We will also refer to the subsets $\ccalS_m^{\text{odd}}, \ccalS_m^{\text{even}}$ for $m\in\{0,1\}$ to denote every odd (even) player from the set $\ccalS_m$ along the line. In the following analysis, we consider six payoff constant cases that ensure an exhaustive analysis for the line, and in the next section, the ring network. First, we consider an odd number of players on the line network.

\begin{proposition}\label{static_odd_line}
	Depending on the utility constants $c_0$ and $c_1$, the optimal control set for the static MPCAC problem for a line network with $n$ odd  is
	\begin{itemize}
		\item[a)] If $1 > 2c_0$ and $1 > 2c_1$, then $\ccalX = \ccalS_1$, $|\ccalX| = \lfloor n/2 \rfloor$, and $\delta_{\ccalS_1} = 0$.
		\item[b)] If $1 > 2c_0$ and $2c_1 > 1 > c_1$, then $\ccalX = \varnothing$.
		\item[c)] If $1 > 2c_0$ and $c_1 > 1$, then $\ccalX = \varnothing$.
		\item[d)] If $2c_0 > 1 > c_0$ and $2c_1 > 1 > c_1$, then $\ccalX = \ccalS_1^{\text{even}}, \{\delta_i\}_{i\in\ccalX} = 0$, and $|\ccalX| = \lfloor n/4 \rfloor$.
		\item[e)] If $c_0 > 1$ and $2c_1 > 1 > c_1$, then $|\ccalX| = 1$ where $\ccalX$ is any one player $i$, with $\delta_i = 0$ ($1$) if $i\in\ccalS_0$ ($\ccalS_1$).
		\item[f)] If $c_0 > 1$ and $c_1 > 1$, then $|\ccalX| = \lceil (n-1)/4 \rceil$, where  $\ccalX =  \ccalS_0^{\text{even}}$, with $\delta_{\ccalX} = 1$.
	\end{itemize}
\end{proposition}
\begin{myproof}
	\begin{itemize}
		\item[a)] The game is dominance solvable with the unique Nash equilibrium $a_i = 1$ for all $i\in\ccalN$. The minimum vertex cover is $\ccalX =\ccalS_1$, and forcing action 0 induces the solution.
		\item[b)] The game is dominance solvable with the unique Nash equilibrium $a_{\ccalS_0} = 1$, $a_{\ccalS_1} = 0$.
		\item[c)] Same proof as (b).
		\item[d)] Consider controlling $\ccalS_1$ players spaced three nodes apart along the line with action 0. These consist of the even-indexed $\ccalS_1$ players. In two time steps, the nodes in between will converge to the equilibrium action inducing maximum anti-coordination. The smallest number of nodes to control in this manner is $ \lfloor n/4 \rfloor$. Sparser placement of control results in undecided players.
		\item[e)] If $\ccalX$ selects a single $i \in \ccalS_1$ to play $a_i=1$, its $\ccalS_0$ neighbors decide on action 0 ($1 < c_0$). At the next iteration, the next $\ccalS_1$ players will decide on action 1 ($1 > c_1$). This cascades down the entire line, resulting in maximum anti-coordination. The same holds if $\ccalX$ selected a single $i \in\ccalS_0$  to play $a_i = 0$.
		\item[f)] In a similar manner as case (d), $\ccalX$ can be chosen to be every other $\ccalS_0$ player along the line network, forcing with action 1 to cause two-hop cascades along the network in two time steps.
	\end{itemize}
\end{myproof}

\begin{proposition}\label{dynamic_odd_line}
	Depending on the utility constants $c_0$ and $c_1$, the optimal control policy $(\{\ccalX^t\},\Delta)$ for the dynamic MPCAC problem on a line network with $n$ odd  is
	\begin{itemize}
		\item[a)] If $1 > 2c_0$ and $1 > 2c_1$, then 
		\begin{equation}
			\begin{cases}\ccalX^t =\varnothing \ &\text{if } t <n \\ \ccalX^t =\ccalS_1, \{\delta_i^t\}_{i\in \ccalS} = 0 &\text{if } t \geq n  \end{cases}
		\end{equation}
		which gives the optimal cost $C^* =  \lfloor n/2 \rfloor$.
		\item[b)] If $1 > 2c_0$ and $2c_1 > 1 > c_1$, then $\ccalX^t = \varnothing$ for all $t \geq 0$.
		\item[c)] If $1 > 2c_0$ and $c_1 > 1$, then $\ccalX^t = \varnothing$ for all $t \geq 0$.
		\item[d)] If $2c_0 > 1 > c_0$ and $2c_1 > 1 > c_1$, then
		\begin{equation}
			\begin{cases}\ccalX^t = \ccalS_1^{\text{even}}, \delta_{\ccalX^t}^t = 0 \ &\text{if } t =0,1 \\ \ccalX^t =\varnothing &\text{if } t > 1  \end{cases}
		\end{equation}
		which gives the optimal cost $C^* =  2\lfloor n/4 \rfloor/n$.
		\item[e)] If $c_0 > 1$ and $2c_1 > 1 > c_1$, then
		\begin{equation}
			\begin{cases}\ccalX^t = \{i\}, \delta_i^t = s_i \ &\text{if } t =0,1 \\ \ccalX^t =\varnothing &\text{if } t > 1  \end{cases}
		\end{equation}
		for any $i \in \ccalN$. This gives the optimal cost $C^* =  2/n$.
		\item[f)] Suppose $c_0 > 1$ and $c_1 > 1$.  Then 
		\begin{equation}
			\begin{cases}\ccalX^t = \ccalS_0^{\text{even}}, \delta_{\ccalX^t}^t = 1 \ &\text{if } t =0,1 \\ \ccalX^t =\varnothing &\text{if } t > 1  \end{cases}
		\end{equation}
		 gives the optimal cost $C^* =  2\lceil (n-1)/4 \rceil/n$.
	\end{itemize}
\end{proposition}
\begin{myproof}
	\begin{itemize}
		\item[a)] The game is dominance solvable, and converges at $t=1$ to $a = (1,\ldots,1)$. By Lemma \ref{thm_vertex_cover}, the optimal policy selects the minimum vertex cover of the line graph after the $n$ time-step horizon.
		\item[b)]  The game is dominance solvable, and converges at $t=2$ to the unique Nash equilibrium $a$ s.t. $a_i = 1 \ \forall i\in\ccalS_0$ and $a_i = 0 \ \forall i\in\ccalS_1$, achieving maximum anti-coordination.  Lemma \ref{thm_dominance_solvable} asserts the optimal policy is the empty set.
		\item[c)] Same proof as (b).
		\item[d)] The action profile $a_{\ccalS_0} = 1$, $a_{\ccalS_1} = 0$ is the only Nash equilibrium achieving maximum anti-coordination. By Lemma \ref{thm_anti_coordination_equilibria}, the optimal policy induces convergence to this equilibrium by $t=n$. Only the endpoint players $\{1,n\}\in\ccalS_0$ have action 1 as a dominant strategy $(1 > c_0)$, and all other players will remain undecided in the absence of control. Hence, the policy that gives convergence to $a$ with minimal cost selects every even $\ccalS_1$ players to play 0 for two time steps, given in the Proposition statement d).
		\item[e)] The game is not dominance solvable, but there are two Nash equilibria $a^*$, $\hat a$ giving maximum anti-coordination, where $(a_{\ccalS_0}^* = 0$, $a_{\ccalS_1}^* = 1)$ and $(\hat{a}_{\ccalS_0} = 1$, $\hat{a}_{\ccalS_1} = 0)$. By Lemma \ref{thm_anti_coordination_equilibria}, the optimal policy induces convergence to one of these equilibria by $t=n$. One can check that a dynamic policy incurring total cost $2/n$ (controlling a single node for two time-steps) induces convergence to $a^*$, and any dynamic policy inducing convergence to $\hat a$ has cost strictly greater than $2/n$ because a player in $i\in\ccalS_1$ needs $a_{i-1}^{t-1} = a_{i+1}^{t-1} = 1$ in order to play the equilibrium action $a_i^t = 0$ (due to $1<2c_1$). Such a convergence would require controlling more than a single node for two time steps.
		\item[f)] The two Nash equilibria $a^*$, $\hat a$ give maximum anti-coordination, so by Lemma \ref{thm_anti_coordination_equilibria}, one of these are selected in optimality. Selecting $\hat a$ gives the minimal cost when every odd $\ccalS_0$ player is controlled with action 1 for two time-steps, which causes the players in between these selections to decide on the equilibrium actions. Similarly, selecting $a^*$ also gives the minimal cost when every odd $\ccalS_1$ player, including the last one, is controlled with action 1 for two time-steps.
	\end{itemize}
\end{myproof}

When considering even number of players $n$ in the line network, we obtain similar results. For brevity, we present the results but omit the proofs because they follow similar arguments as Proposition \ref{dynamic_odd_line}. Here, the type configuration in the network also alternates s.t. $i \in \ccalS_0$ if $i$ is odd, and $s_i \in \ccalS_1$ if $i$ is even.

\begin{proposition}\label{even_line_static}
	Depending on the utility constants $c_0$ and $c_1$, the optimal solutions to static MPCAC for a line network with $n$ even is
	\begin{itemize}
		\item[a)] If $1 > 2c_0$ and $1 > 2c_1$, then $\ccalX = \ccalS_0$ or $\ccalS_1$ with $\delta_{\ccalX} = 0$. Then, $|\ccalX| = n/2$.
		\item[b)] If $1 > 2c_0$ and $2c_1 > 1 > c_1$, then $\ccalX = \{n\}$ with $\delta_n = 0$.
		\item[c)] If $1 > 2c_0$ and $c_1 > 1$, then $\ccalX = \varnothing$. 
		\item[d)] If $2c_0 > 1 > c_0$ and $2c_1 > 1 > c_1$, then $\ccalX = \ccalS_1^{\text{odd}}$ with $\delta_{\ccalX} = 0$. Then $|\ccalX| = \lceil n/4 \rceil$.
		\item[e)] If $c_0 > 1$ and $2c_1 > 1 > c_1$, then $\ccalX = \varnothing$. 
		\item[f)] If $c_0 > 1$ and $c_1 > 1$, then $\ccalX = \ccalS_1^{\text{odd}}$ with $\delta_{\ccalX} = 1$. Then $|\ccalX| = \lceil n/4 \rceil$.
	\end{itemize}
\end{proposition}

\begin{proposition}\label{even_line_dynamic}
	Depending on the utility constants $c_0$ and $c_1$, the optimal solutions to the dynamic MPCAC problem for a line network with $n$ even is
	\begin{itemize}
		\item[a)] If $1 > 2c_0$ and $1 > 2c_1$, then 
		\begin{equation}
			\begin{cases}\ccalX^t =\varnothing \ &\text{if } t <n \\ \ccalX^t = \ccalS_0 \ \text{or } \ccalS_1, \{\delta_i^t\}_{i\in \ccalS} = 0 &\text{if } t \geq n  \end{cases}
		\end{equation}
		which gives the optimal cost $C^* =   n/2 $.
		\item[b)] If $1 > 2c_0$ and $2c_1 > 1 > c_1$, then
		\begin{equation}
			\begin{cases}\ccalX^t =\varnothing \ &\text{if } t <n \\ \ccalX^t = \{n\}, \delta_n^t = 0 &\text{if } t \geq n  \end{cases}
		\end{equation}
		which gives the optimal cost $C^* =   1 $.
		\item[c)] If $1 > 2c_0$ and $c_1 > 1$, then $\ccalX^t = \varnothing$ for all $t \geq 0$.
		\item[d)] If $2c_0 > 1 > c_0$ and $2c_1 > 1 > c_1$, then
		\begin{equation}
			\begin{cases}\ccalX^t =\ccalS_1^{\text{odd}} \cup \{n\}, \delta_{\ccalX^t}^t = 0 \ &\text{if } t =0,1 \\ \ccalX^t =\varnothing &\text{if } t > 1  \end{cases}
		\end{equation}
		which gives the optimal cost $C^* = 2\lceil n/4 \rceil/n$.
		\item[e)] If $c_0 > 1$ and $2c_1 > 1 > c_1$, then
		\begin{equation}
			\begin{cases}\ccalX^t = \{i\}, \delta_i^t = s_i \ &\text{if } t =0,1 \\ \ccalX^t =\varnothing &\text{if } t > 1  \end{cases}
		\end{equation}
		for any $i \in \ccalN$. This gives the optimal cost $C^* =  2/n$.
		\item[f)] Suppose $c_0 > 1$ and $c_1 > 1$. Then
		\begin{equation}
			\begin{cases}\ccalX^t = \ccalS_1^{\text{odd}}, \delta_{\ccalX^t}^t = 1 \ &\text{if } t =0,1 \\ \ccalX^t =\varnothing &\text{if } t > 1  \end{cases}
		\end{equation}
		 gives the optimal cost $C^* =  2\lceil n/4 \rceil/n$.
	\end{itemize}
\end{proposition}

\subsection{Ring network}
Here, we consider a ring network with alternating type configuration and where $n$ is even. For convention, we index the players in the same fashion as the line network above, and alternate types between $\ccalS_0, \ccalS_1$ - i.e., odd (even) indices are $\ccalS_0$ ($\ccalS_1$) players.
\begin{proposition}\label{static_even_ring}
	Depending on the utility constants $c_0$ and $c_1$, the optimal solutions to static MPCAC for a ring  network with $n$ even  is
	\begin{itemize}
		\item[a)] If $1 > 2c_0$ and $1 > 2c_1$, $\ccalX = \ccalS_0$ or $\ccalS_1$, and $|\ccalX| = n/2$.
		\item[b)] If $1 > 2c_0$ and $2c_1 > 1 > c_1$, $\ccalX = \varnothing$.
		\item[c)] If $1 > 2c_0$ and $c_1 > 1$, $\ccalX = \varnothing$.
		\item[d)] If $2c_0 > 1 > c_0$ and $2c_1 > 1 > c_1$,  $\ccalX = \ccalS_0^{\text{odd}}$ or $\ccalS_1^{\text{odd}}$ with $\delta_{\ccalX} = 1$. Then, $|\ccalX| = \lceil n/4 \rceil$.
		\item[e)] If $c_0 > 1$ and $2c_1 > 1 > c_1$,  $\ccalX = \{i\}$ with $\delta_i = s_i$ for any $i\in\ccalN$. Then  $|\ccalX| = 1$. 
		\item[f)] If $c_0 > 1$ and $c_1 > 1$,  either $\ccalX =\ccalS_0^{\text{odd}}$ or $\ccalX =\ccalS_1^{\text{odd}}$ with $\delta_{\ccalX} = 1$. Then $|\ccalX| = \lceil n/4 \rceil$.
	\end{itemize}
\end{proposition}
\begin{myproof}
	\begin{itemize}
		\item[a)] The game is dominance solvable with all nodes playing action 1. Hence, influence of either all type-0 or type-1 nodes gives the solution.
		\item[b)] The game is dominance solvable, which converges to maximum anti-coordination (type-0 playing 1, type-1 playing 0) without influencing any nodes.
		\item[c)] Same argument as (b).
		\item[d)] The learning algorithm produces no decided nodes. Controlling a single node to play 0 results in the algorithm converging in the next iteration, with its two neighbors playing 1. Hence, by controlling every other node of one type (every 4 along the ring) to play 0, the algorithm will converge to maximum anti-coordination in two iterations.
		\item[e)] The algorithm produces no decided nodes. If a single type-0 node is forced to play 0, decisions will be cascaded along the ring resulting in maximum anti-coordination. Similarly, one could influence a single type-1 node to play 1.
		\item[f)] Same argument as (d).
	\end{itemize}
\end{myproof}
\begin{proposition}\label{dynamic_even_ring}
	Depending on the utility constants $c_0$ and $c_1$, the optimal solutions to dynamic MPCAC for a ring  network with $n$ even  is
	\begin{itemize}
		\item[a)] If $1 > 2c_0$ and $1 > 2c_1$
		\begin{equation}
			\begin{cases}\ccalX^t =\varnothing \ &\text{if } t <n \\ \ccalX^t = \ccalS_0 \ \text{or } \ccalS_1, \{\delta_i^t\}_{i\in \ccalS} = 0 &\text{if } t \geq n  \end{cases}
		\end{equation}
		The resulting optimal cost is $C^* = n/2$.
		\item[b)] If $1 > 2c_0$ and $2c_1 > 1 > c_1$, $\ccalX^t = \varnothing$ for all $t\geq 0$.
		\item[c)] If $1 > 2c_0$ and $c_1 > 1$, $\ccalX^t = \varnothing$ for all $t\geq 0$.
		\item[d)] If $2c_0 > 1 > c_0$ and $2c_1 > 1 > c_1$,
		\begin{equation}
			\begin{cases}\ccalX^t = \ccalS_1^{\text{odd}}, \delta_{\ccalX^t}^t = 1 \ &\text{if } t =0,1 \\ \ccalX^t =\varnothing &\text{if } t > 1  \end{cases}
		\end{equation}
		The resulting optimal cost is $C^* =  2\lceil n/4 \rceil/n$.
		\item[e)] If $c_0 > 1$ and $2c_1 > 1 > c_1$,
		\begin{equation}
			\begin{cases}\ccalX^t = \{i \}, \delta_{i}^t = s_i \ &\text{if } t =0,1 \\ \ccalX^t =\varnothing &\text{if } t > 1  \end{cases}
		\end{equation}
		The resulting optimal cost is $C^* = 2/n$.
		\item[f)] If $c_0 > 1$ and $c_1 > 1$, then either
		\begin{equation}
			\begin{cases}\ccalX^t = \ccalS_0^{\text{odd}}, \delta_{\ccalX^t}^t = 1 \ &\text{if } t =0,1 \\ \ccalX^t =\varnothing &\text{if } t > 1  \end{cases}
		\end{equation}
		or
		\begin{equation}
			\begin{cases}\ccalX^t = \ccalS_1^{\text{odd}} , \delta_{\ccalX^t}^t = 1 \ &\text{if } t =0,1 \\ \ccalX^t =\varnothing &\text{if } t > 1  \end{cases}
		\end{equation}
		 gives the optimal cost $C^* =  2\lceil n/4 \rceil/n$.
	\end{itemize}
\end{proposition}
\begin{myproof}
	We omit the details of the proof for brevity. The proof utilizes similar arguments from Proposition \ref{dynamic_odd_line} by invoking Lemmas \ref{thm_dominance_solvable}, \ref{thm_anti_coordination_equilibria}, and  \ref{thm_vertex_cover}, as well as the structural arguments from the proof of Proposition \ref{static_even_ring}.
\end{myproof}

\subsection{Discussion}

The optimal control policies vary depending on the payoff constants. If all players decide to play action 1 as a result of the learning process, then we solve a minimum vertex covering problem along the lines of Lemma \ref{thm_vertex_cover}, and the players selected have to be controlled for all times $t\geq n$---case (a) in all of the propositions in this section. If the game is dominance solvable and there are no active anti-coordination links in the equilibrium action profile, the optimal policy is the empty set---see Propositions \ref{thm_star_staticMPCAC}(b), \ref{thm_star_dynamicMPCAC}(b), \ref{static_odd_line}(b,c), \ref{dynamic_odd_line}(b,c), \ref{even_line_static}(b,c) \ref{even_line_dynamic}(b,c), \ref{static_even_ring}(b,c), and \ref{dynamic_even_ring}. If a subset of players remain undecided as a result of the learning algorithm and there are no failed anti-coordination links between the decided players, an optimal policy can be found by leveraging Lemma \ref{thm_anti_coordination_equilibria}. In the benchmark examples, these policies required controlling a subset of the players, causing the undecided players in the other type to decide. In dynamic MPCAC cases, this subset needed to be controlled for two time steps. The neighboring players will decide on actions after the first time step, and the controlled players use their neighbors' decisions to consolidate their (unforced) actions in the second time step. Thus, the control can be lifted and no player will revert from their decided actions. The resultant action profile is a Nash equilibrium action profile that satisfies maximum anti-coordination. Hence, there is no need to make control efforts after time $n$. Consequently, we can control a smaller subset of players and leverage the learning dynamics to create cascades of decision-making---see cases (d-f) in Propositions \ref{static_odd_line} - \ref{dynamic_even_ring}. In the latter scenarios, the policy suggested by Lemma \ref{thm_vertex_cover} is an upper bound of the optimal policy.

%% file: star.tex

\usetikzlibrary{matrix,arrows,decorations.pathmorphing}
\usetikzlibrary{arrows,automata}

{\small
\tikzstyle{terminal} = [circle, draw=black, inner sep=0pt, minimum size=0.5cm]
\tikzstyle{theta} = [circle, draw=white, inner sep=0pt, minimum size=0.5cm]
\tikzstyle{arrow} = [stealth-stealth, thick]

\begin{tikzpicture}[x=0.4cm, y=0.3cm]

\draw
       ++( 0:0) node (terminal 1)  [terminal, fill=red!30,label= 90:{}] {1} ;
        
\draw
       ++( 270:4) node (terminal 2)  [terminal, fill=blue!30, label= 270:{}] {2} 
	++( 0:3) node (terminal 3)  [terminal, fill=blue!30, label= 270:{}] {3}
        ++( 0:3) node (terminal 4)  [terminal, fill=blue!30, label= 270:{}] {4};
                
\draw [arrow] (terminal 3) to [bend right = 0] (terminal 1);
\draw [arrow] (terminal 2) to [bend left = 0] (terminal 1);
\draw [arrow] (terminal 4) to [bend right = 0] (terminal 1);


\end{tikzpicture}
}

%% file: line.tex

\usetikzlibrary{matrix,arrows,decorations.pathmorphing}
\usetikzlibrary{arrows,automata}

{\small
\tikzstyle{terminal} = [circle, draw=black, inner sep=0pt, minimum size=0.5cm]
\tikzstyle{theta} = [circle, draw=white, inner sep=0pt, minimum size=0.5cm]
\tikzstyle{arrow} = [stealth-stealth, thick]

\begin{tikzpicture}[x=0.4cm, y=0.3cm]

\draw
       ++( 0:0) node (terminal 1)  [terminal, fill=blue!30, label= 90:{}] {1} 
       	++( 0:3) node (terminal 2)  [terminal, fill=blue!30, label= 90:{}] {2};
        
\draw
	++( 270:4) node (terminal 3)  [terminal, fill=red!30, label= 270:{}] {3}
        ++( 0:3) node (terminal 4)  [terminal, fill=red!30, label= 270:{}] {4};
                
\draw [arrow] (terminal 3) to [bend right = 0] (terminal 1);
\draw [arrow] (terminal 2) to [bend left = 0] (terminal 3);
\draw [arrow] (terminal 4) to [bend right = 0] (terminal 2);


\end{tikzpicture}
}

%% file: ring.tex

\usetikzlibrary{matrix,arrows,decorations.pathmorphing}
\usetikzlibrary{arrows,automata}

{\small
\tikzstyle{terminal} = [circle, draw=black, inner sep=0pt, minimum size=0.5cm]
\tikzstyle{theta} = [circle, draw=white, inner sep=0pt, minimum size=0.5cm]
\tikzstyle{arrow} = [stealth-stealth, thick]

\begin{tikzpicture}[x=0.4cm, y=0.3cm]

\draw
       ++( 0:0) node (terminal 1)  [terminal, fill=red!30, label= 90:{}] {1} 
       	++( 0:3) node (terminal 2)  [terminal, fill=red!30, label= 90:{}] {2};
        
\draw
	++( 270:4) node (terminal 3)  [terminal, fill=blue!30, label= 270:{}] {3}
        ++( 0:3) node (terminal 4)  [terminal, fill=blue!30, label= 270:{}] {4};
                
\draw [arrow] (terminal 3) to [bend right = 0] (terminal 1);
\draw [arrow] (terminal 2) to [bend left = 0] (terminal 3);
\draw [arrow] (terminal 4) to [bend right = 0] (terminal 2);
\draw [arrow] (terminal 4) to [bend right = 0] (terminal 1);


\end{tikzpicture}
}

%% file: conclusion.tex

We considered the control of players' learning processes in a population to influence the emergent outcome in the context of anti-coordination network games. With the goal to promote maximum anti-coordination with minimum effort, we developed computationally tractable methods that determine when to control which players, and how to control them. An algorithm that sequentially selected players according to their influence in promoting anti-coordination in the future performed well in random networks with arbitrary population sizes.

%% file: appendix.tex

\subsection{Proof of Theorem \ref{local_algorithm_convergence}}

We show that Algorithm \ref{local_alg} is equivalent to the iterated elimination of strictly dominated strategies in Definition \ref{def:elimination_strict}. This equivalence yields convergence to Nash equilibrium if the game is dominance solvable. Otherwise, it yields elimination of all the strictly dominated actions. 

We prove by induction. Given $a_i^0 = \vec{\epsilon}$, at time $k=1$ each player that selects action 1 ($a_i^1$) by \eqref{worst_case_good_response} eliminates all the other actions in $[0,1)$. To see this consider the best response of player $i$ given in \eqref{best_response_specific} where $BR_i(a_{\ccalN_i}) \geq BR_i(\lceil a_{\ccalN_i}^0 \rceil)$ for all $a_{\ccalN_i} \in [0,1]^{|\ccalN_i|}$. Further note that $\lceil a_{\ccalN_i}^0 \rceil = \bbone_{|\ccalN_i|}$. If $BR_i(\bbone_{|\ccalN_i|}) = 1$ then it is best to play action 1 against all possible actions of neighboring players by the previous inequality. Hence, by \eqref{eq:strictly_dominated_definition}, all actions in $[0,1)$ are strictly dominated by  $a_i^1 = 1$.

Consider now time $k$ given action profile $a^{k-1}$ where players that decided $a^{k-1}_i \in \{0,1\}$ have eliminated rest of their actions. Define the not strictly dominated action space of each player $i\in\ccalN$ as  $A_i^{k-1} = \{1\}$ if $a^{k-1}_i=1$, $A_i^{k-1} = \{0\}$ if $a^{k-1}_i=0$, and $A_i^{k-1} = [0,1]$ if $a^{k-1}_i=\epsilon$ (recall the notation in Definiton \eqref{def:elimination_strict}). For any player $i$,  we have $BR_i(a_{\ccalN_i}) \geq BR_i(\lceil a_{\ccalN_i}^{k-1} \rceil)$  and $BR_i(a_{\ccalN_i}) \leq BR_i(\lfloor a_{\ccalN_i}^{k-1} \rfloor)$ for any $a_{\ccalN_i}\in A_{\ccalN_i}^{k-1}$. Using the first inequality, if $BR_i(\lceil a_{\ccalN_i}^{k-1} \rceil)=1$ then it is best to play action 1 against all possible non-dominated actions of neighboring players. That is, action 1 dominates all possible actions of player $i$. Using the second inequality, if $BR_i(\lfloor a_{\ccalN_i}^{k-1} \rfloor) = 0$ then it is best to play action 0 against all possible non-dominated actions of neighboring players. That is, action 0 dominates all possible actions of player $i$.

Given the definition of the best response in \eqref{best_response_specific}, eliminating all the actions except 1 if $BR_i(\lceil a_{\ccalN_i}^{k-1} \rceil)=1$ is equivalent to the condition given in \eqref{worst_case_good_response}. Similarly, eliminating all the actions except 0 if $BR_i(\lfloor a_{\ccalN_i}^{k-1} \rfloor)$ is equivalent to the condition given in \eqref{best_case_bad_response}. Combined with the above argument, given the action profile at time $k-1$ ($a^{k-1}$), one step of Algorithm \ref{local_alg} yields an action profile $a^k$ where decided players eliminate all strictly dominated actions given the not strictly dominated action space $A^{k-1}$. Since at time $k=1$ the update eliminated strictly dominated strategies, Algorithm \ref{local_alg} is equivalent to the iterated elimination process given in Definition \ref{def:elimination_strict} by induction. 

Given this equivalence, if the game is dominance solvable, the algorithm converges to the unique Nash equilibrium. Otherwise, all decided players eliminate all of the actions except the action they selected, and undecided players cannot eliminate any actions from the their initial action space $[0,1]$. 

Next, we prove convergence in $n$ time steps. Suppose at time step $k$ given $a^{k-1}$, there does not exist a player that switches from being undecided, i.e., $a_i^{k-1}=\epsilon$, to being decided, i.e., $a_i^{k} = \{0,1\}$. That is, if $a_i^{k}=a_i^{k-1}$ then $a_i^{k+1} = a_i^k$. Further, if a player is decided, it cannot change its action because all the other possible actions are dominated, that is, if $a_i^k=\{0,1\}$, then $a_i^{k+1} = a_i^k$. Given these two observations, at least one player has to switch to being decided at time $k-1$, in order for at least one player to become decided at time $k$ given that it was undecided at time $k-1$. There can at most be $n$ instances of switching from being undecided to being decided which is the case when the game is dominance solvable. Further there needs to be at least one switching happening at each time step for the updates to continue. Hence, the algorithm converges in at most $n$ steps. 

\subsection{Proof of Lemma \ref{thm_dominance_solvable}}
Dominance solvability implies Algorithm \ref{local_alg} converges to a unique Nash equilibrium $a^{\text{NE}}$ defined in \eqref{Nash_definition}. This means the second constraint of dynamic MPCAC is satisfied by time $n$. There are two possible Nash equilibrium action profiles: $a^{NE}$ satisfies maximum anti-coordination constraint (first constraint) in \eqref{eqn_objective}, or $a^{\text{NE}}$ does not satisfy it. If $a^{\text{NE}}$ satisfies the maximum anti-coordination constraint, then the optimal control profile that minimizes the objective is the empty set ($\ccalX^t_* = \emptyset$). If $a^{\text{NE}}$ does not satisfy maximum anti-coordination constraint, we need to control a non-empty set of players to achieve maximum anti-coordination by time $n$. Assume now the optimal control profile after time $n$ is empty, i.e., $\ccalX^t_* = \emptyset$ for $t>n$. Then because the game is dominance solvable, we have $a^{2n}=\Phi_n(a^n) = a^{\text{NE}}$ where $\Phi_k(\cdot)$ is defined in \eqref{eq_action_profile_k}. That is, by time $t=2n$, players will revert to $a^{\text{NE}}$ which violates the first constraint in \eqref{eqn_objective}. Hence, a control profile $\ccalX^t_* = \emptyset$ for $t>n$ cannot be optimal because it is not feasible when $a^{\text{NE}}$ is not feasible.

\subsection{Proof of Lemma \ref{thm_anti_coordination_equilibria}}

Suppose the optimal policy $\Pi^*$ to dynamic MPCAC in \eqref{eqn_objective} is such that it does not converge to an equilibrium of the game $a^*$ that achieves maximum anti-coordination, i.e., $a_i^* + a_j^* \leq 1$ for all $(i,j)\in\ccalE_B$. It means there must  at least be one player controlled after $t\geq n$ to achieve maximum anti-coordination for all times $t\geq n$. That is, in the best case scenario we have to control a single player after $t\geq n$. Further, there must at least be one player that is controlled for one time step before time $n$ to achieve anti-coordination at time $n$. Combining the cost for control before and after time $n$, the optimal value attained by $\Pi^*$ will at least be $1/n+1$. 

If there exists an equilibrium $a^*$ that achieves maximum anti-coordination, then Algorithm \ref{local_alg} cannot converge to an action profile $a^n$ with an active link, i.e., a link $(i,j)\in \ccalE_B$ where $a_i^n + a_j^n > 1$, because it would imply the equilibrium action profile $a^*$ is strictly dominated by an action profile, $a^n$. This means that Algorithm \ref{local_alg} can only eliminate possibly active links. Hence, the worst case scenario in terms of cost of control is when all players remain undecided as a result of Algorithm \ref{local_alg}, i.e., $a^n = \vec\epsilon$. In this case, a feasible solution is one where we control the players in a given type ($\ccalX^t = \{i: s_i =0\}$ or $\ccalX^t =\{i:s_i =1\}$) for two time steps, e.g., for $t=0,1$. This causes the players in the other type $\ccalN\setminus\ccalX^0$ to play action 1 by \eqref{eq_dynamics} at $t=1$. That is, $a_i^0 = 0$ if $i\in \ccalX^0$, and $a_i^0 = 1$ if $i\in\ccalN\setminus \ccalX^0$. In step $t=1$, we continue to implement the control by $\ccalX^1$, then the players in $\ccalX^1$ take action 0 by \eqref{eq_dynamics}, i.e., $a_i^1=x_i^1 \delta_i^1=0$ for $i\in\ccalX^1$, and $a_i^1 = y_i^1=\Phi(a^0)= 1$ if $i\in\ccalN\setminus \ccalX^1$. When we remove the control of players in $\ccalX^t$ for $t>2$, players do not change their actions because the resultant action profile is an equilibrium. If we select the type with the lowest number of players, then our cost for this control policy is at most $\lfloor n/2\rfloor$ for one step. Hence, the objective value in \eqref{eqn_objective} is at most $(|\ccalX^0|+|\ccalX^1|)/n=1$. This policy is an upper bound on the optimal policy. Consequently, $\Pi^*$ that achieves a cost of $1/n+1$ cannot be optimal. 

\subsection{Proof of Corollary \ref{thm_optimal_vertex_cover}}

When all players are decided in a single time step, all players decide on taking action 1 using \eqref{worst_case_good_response}, that is, $a = \bbone_{n\times1}$ where $= \bbone_{n\times1}$ is an $n\times1$ vector with all elements equal to one.

\begin{claim} \label{cl_optimal_empty}
$\ccalX^t_* =\emptyset$ for $t<n$.
\end{claim}
\begin{myproof}
Suppose there exists an optimal policy $\Pi$  such that $\ccalX^{\bar t}_* \neq\emptyset$ for $\bar t<n$. Given the controlled action profile at time $\bar t$, $a^{\bar t}$, we have the uncontrolled action profile at time $t+1$ as $y^{\bar t+1} = \bbone$. Hence, the uncontrolled action profile at time $n$ is given by $\bbone$. As a result, a policy where $\ccalX^t =\emptyset$ for $t<n$ and $\ccalX^t =\ccalX^t_*$ for $t\geq n$ would be feasible and would incur a smaller cost than the policy $\Pi$ by an amount $\frac{1}{n}\sum_{t=1}^{n-1} |\ccalX^t_*|$.
\end{myproof}

We continue with the proof of Corollary \ref{thm_optimal_vertex_cover}. By Claim \ref{cl_optimal_empty}, no control action is taken until time $n$. At time $n$, we have $y^n = \bbone_{n\times1}$. Given $y^n$, the optimal control policy at time $n$ is given by the following single time-step optimization,
\begin{align}
\min_{\ccalX} &  |\ccalX| \label{eq_cardinality}\\
\text{s.t. } &a_i+a_j \leq 1 \quad \forall (i,j) \in \ccalE_B\\
& a_i =  1-x_i \quad \forall i\in\ccalN\\
& a_i \in \{0,1\}. 
\end{align}
We note that if a player is controlled, $x_i=1$, then it must be that $\delta_i = 0$ because $a_i^n= 1$ for all $i$. Hence, we have the second constraint from the controlled dynamics, $(1-x_i) a_i^n + \delta_i x_i = 1-x_i$. Define the dynamic control policy $\bar\ccalX$ where $\bar\ccalX^t=\emptyset$ for $t<n$, and we implement $\bar\ccalX^n$, which is the solution to the above optimization problem, for $t\geq n$.  

Suppose there exists $\tilde\ccalX$ that achieves a lower cost than $\bar\ccalX$ in the dynamic MPCAC problem. Given the first constraint above it is guaranteed that the controlled dynamics $\Phi_n(\vec{\epsilon},\bar\ccalX,\Delta)$ yield an action profile at time $n$ that achieves anti-coordination.

By Claim \ref{cl_optimal_empty}, it must be that $\tilde\ccalX^t =\emptyset$ for $t<n$.  Note that there cannot exist policy at time $n$ such that $|\tilde\ccalX^n|< |\bar\ccalX^n|$ because $\bar\ccalX^n$ is an optimal solution of \eqref{eq_cardinality}. If $|\tilde\ccalX^n|> |\bar\ccalX^n|$ then we can use $\bar\ccalX$ to obtain a smaller cost ($\sum_{t=1}^{n} |\tilde\ccalX^t|>\sum_{t=1}^{n} |\bar\ccalX^t|=|\bar\ccalX^n|$) for the first $n$ time steps. Further, the uncontrolled action profile at time $n+1$ is given by $y^{n+1}= \Phi_1(a^n)=\bbone_{n\times1}$ where controlled action of player $i$ is given by $ a_i^n = 1-x_i^*$ for $x_i^* \in \bar\ccalX^n$. Hence, there cannot exist a control policy $\tilde\ccalX^t$ such that $|\tilde\ccalX^t| < |\bar\ccalX^t|$ for $t>n$ by the same reasoning as above. Combining the above findings, we have the dynamic MPCAC objective for the control policy $\tilde\ccalX$ as $\frac{1}{n} |\tilde\ccalX^n|+ \lim_{T' \rightarrow \infty} \frac{1}{T'}\sum_{t=n+ 1 }^{T'} |\tilde\ccalX^t| >(1/n+1)|\bar\ccalX^n|$. This is a contradiction.

Substituting $a_i = 1-x_i$ in the constraint $a_i + a_j\leq 1$, we get $x_i + x_j \geq 1$ when $x_i \in\{0,1\}$ and $x_j \in\{0,1\}$. Note that $\ccalE_B = \ccalE^n$ and $\ccalN^n = \ccalN$ given $a_i^n=1$. Hence, the last constraint in \eqref{eq_objective_vertex_cover} does not exist when $a^n = \bbone$. This shows that the optimization in \eqref{eq_cardinality} is equivalent to the optimization problem \eqref{eq_objective_vertex_cover} when $a^n =\bbone$.